\documentclass[a4paper,aps,superscriptaddress,floatfix,nofootinbib,twocolumn,longbibliography]{revtex4-1}

\usepackage{bbold}
\usepackage{bbm}
\usepackage[pdftex]{graphicx}
\usepackage{latexsym,amsmath,verbatim,amssymb,txfonts}
\usepackage{color}
\usepackage{rotating}
\usepackage{verbatim}
\usepackage{multirow}
\usepackage[english]{babel}
\usepackage{comment}
\usepackage{braket}
\usepackage{amsmath}
\usepackage{color}

\newcommand{\change}[1]{ #1}

\begin{document}

\title{Decoding communities in networks}

\author{Filippo Radicchi}
\affiliation{Center for Complex Networks and Systems Research, School
  of Informatics, Computing, and Engineering, Indiana University, Bloomington,
  Indiana 47408, USA}
\email{filiradi@indiana.edu}

\begin{abstract}
According to a recent information-theoretical proposal, the 
problem of defining and identifying communities in networks
can be interpreted as a classical 
communication task over a noisy
channel: memberships of nodes are information 
bits erased by the channel, 
edges and non-edges in the network are parity 
bits introduced by the encoder
but degraded through the channel, and a 
community identification
algorithm is a decoder. The interpretation
is perfectly equivalent to the one at the basis of
well-known statistical inference algorithms for community
detection. The only difference in the interpretation is that a noisy channel replaces 
a stochastic network model. However, the different perspective
gives the opportunity to take advantage of the rich set of 
tools of coding theory to generate novel 
insights on the problem of community detection.  In this paper, we 
illustrate two main 
applications of standard coding-theoretical methods 
to community detection.
First, we leverage a state-of-the-art
decoding technique to generate a family of quasi-optimal
community detection algorithms. Second and more important, 
we show that the  Shannon's noisy-channel coding theorem can 
be invoked to establish a lower bound, here named as decodability
bound,  for the maximum amount of 
noise tolerable by an ideal decoder to achieve perfect detection of
communities. When computed for well-established synthetic benchmarks, 
the decodability bound explains accurately
the performance achieved by the best 
community detection algorithms existing on the market, telling us that
only little room for their improvement is still potentially left. 
\end{abstract}

\maketitle

\section{Introduction}
Real networks are often assumed to be organized in clusters or
communities~\cite{fortunato2010community}.
A community is naively defined as
a subgroup of nodes with a density of internal connections 
larger than the density of external links.
Most of the research in the area has focused
on the development of algorithms aimed at detecting such 
objects. The philosophy of the approaches 
considered so far varies widely, with
methods that rely on
heuristics~\cite{girvan2002community, radicchi2004defining}, spectral
properties of operators~\cite{newman2013spectral}, and 
optimization of quality functions~\cite{newman2004finding,
  reichardt2004detecting, rosvall2007information, rosvall2008maps}, 
just to mention a few of them.
Principled approaches, 
as those relying on generative network models
~\cite{decelle2011inference,
karrer2011stochastic, peixoto2014hierarchical,
peixoto2013parsimonious, peixoto2017bayesian, moore2017computer}, 
provide not only practical algorithms, 
but also a solid notion of a community. 
In this respect, they allow to 
generate insights on the  problem of identification of communities
in networks, as for example establishing the existence of a universal
limitation affecting all community detection algorithms
~\cite{decelle2011inference,
  nadakuditi2012graph, krzakala2013spectral,
  radicchi2013detectability, radicchi2014paradox}.
The limitation refers to the performance of 
a perfect algorithm in stochastic network models with planted
community structure, and consists in the existence of a
maximum level of fuziness, generally named as detectability threshold, that can 
be tolerated by the algorithm to be able to detect, in the limit of
infinite network sizes, a non-vanishing portion of the true community structure.
As statistical inference approaches
rely on stochastic network models~\cite{decelle2011inference,
karrer2011stochastic, peixoto2014hierarchical,
peixoto2013parsimonious, peixoto2017bayesian, moore2017computer}, the detectability threshold of
these models provides an indication of the 
parameter ranges where community detection algorithms
are expected to be useful. 
\change{Although rigorous conditions for the
establishment of the regime of detectability 
has been studied also
for finite-size networks~\cite{PhysRevE.95.062304}, the 
notion of detectability is
much less useful in another important application of
stochastic block models, that is the numerical validation
of community detection 
algorithms~\cite{girvan2002community, danon2005comparing, 
lancichinetti2008benchmark, lancichinetti2009community}.
In this type of application in fact, the focus is
\change{not only on network models with small/medium size}, but,
 more importantly, } 
on the regime of exact 
recovery of the planted community structure.

\begin{figure*}[!htb]
\begin{center}
\includegraphics[width=0.95\textwidth]{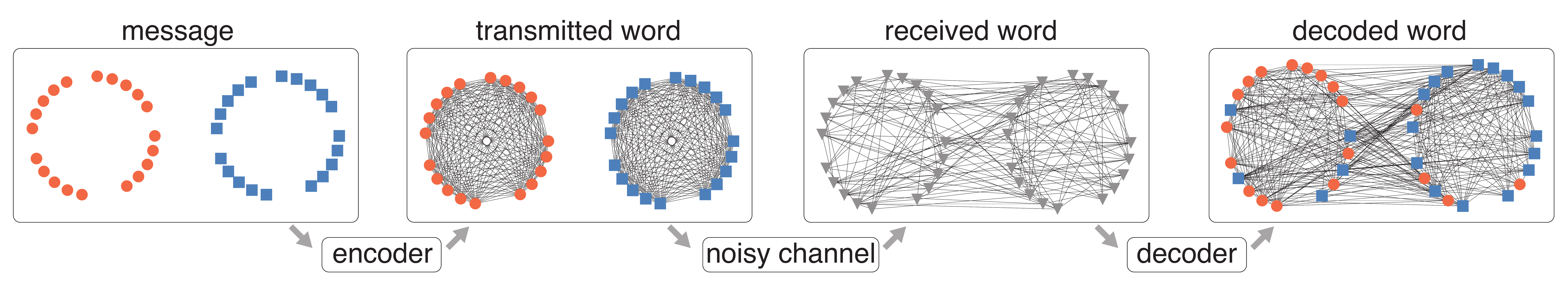}
\end{center}
\caption{(Color online) Definition and detection of communities in networks  
as a communication process~\cite{abbe2016exact, abbe2015community, 
abbe2017community}.
The message generated from the source contains information about
the group assignments of the various nodes. The message is encoded 
in a network structure where all nodes belonging to the same group are 
connected one with the other, but no connections are present between any pair
of nodes belonging to different groups. The codeword is transmitted
along a noisy channel. The effect of the channel is twofold: it erases all
information about group assignments, and deteriorates the network
structure by adding and removing edges. The received word is
decoded to reconstruct, even if only partially, the 
transmitted message.
}
\label{fig:1}
\end{figure*}

The formal establishment of the regimes of partial and exact recovery 
in the stochastic block model has been the subject of a series
of recent publications in coding theory~\cite{mossel2013proof, abbe2016exact, abbe2015community, 
abbe2017community}. In these papers, 
the problem of defining and identifying communities is interpreted as a classical
communication process (Fig.~\ref{fig:1}), analogous to the one considered by 
Shannon~\cite{shannon}: group assignments of the 
nodes in the network represent a message that is first 
encoded, then
transmitted along a noisy channel, and ultimately 
decoded. 
 It is important to remark that
the noisy-channel interpretation of the problem of community detection
coincides with the one at the basis of
statistical inference approaches~\cite{decelle2011inference,
karrer2011stochastic, peixoto2014hierarchical,
peixoto2013parsimonious, peixoto2017bayesian, moore2017computer}.
The only difference is the angle from which one looks at it. 
Instead of interpreting  a stochastic model as the generator of noisy edges,
the stochastic model is seen as a source of noise that disrupts
a network with community structure otherwise unambiguous.
In this respect, edges in the network are not regarded 
as entities that define communities rather as redundant but altered
information that is added to pre-existing information
on node memberships. Interpreting the task of detecting 
communities in graphs
as a decoding process of a noisy signal has the great
advantage
to lead to rigorous mathematical statements, valid in the limit
of infinitely large graphs, regarding
the identification of the regimes of partial and exact reconstruction
of the planted community structure of the stochastic block 
model, regimes that are entirely determined by the amount of noise 
that characterizes the network model. Partial recovery corresponds to
the aforementioned detectability~\cite{mossel2013proof}. 
The determination of the
exact recovery threshold is instead a completely novel result,
able to provide a precise indication of the range of parameter values of
a stochastic block model where perfect recovery of the planted
community structure is allowed~\cite{abbe2016exact}. The computation of the
exact recovery threshold can be performed in general
stochastic block models~\cite{abbe2015community}. However, the
mathematical result 
is derived in the limit of infinitely large systems. 
On finite systems, we
expect the threshold to delineate conservative
regimes of performance. 
How much is the threshold value underestimating the potential
of algorithms on mid-sized networks as those used in
typical numerical experiments~\cite{danon2005comparing, 
lancichinetti2009community}? Are there more predictive approximations
for finite-size networks?  Please note that this information
is of fundamental importance. Without a theoretical baseline, we can
assess the performance of algorithms only in pairwise comparisons.
This means that
we don't have the ability to
judge their full potential, and
 determine if there is still room for improvement.

The current paper aims at filling this gap. We provide 
\change{ numerical } evidence that the
exact recovery threshold~\cite{abbe2016exact, abbe2015community, 
abbe2017community} generates predictions that
seem to be not very accurate in the 
description of the performance of algorithms in finite-size
networks~\cite{danon2005comparing, lancichinetti2009community}.
\change{Specifically, we provide numerical evidence
that existing community detection algorithms
are able to achieve perfect recovery of communities 
in small/medium stochastic block models well beyond 
the exact recovery threshold~\cite{danon2005comparing, lancichinetti2009community}.}
We show, however, that the Shannon's noisy-channel coding
theorem~\cite{shannon} allows us to establish a 
less restrictive approximation for the regime of exact recovery
which describes particularly well the maximum level of fuziness tolerated by
algorithms to achieve perfect detection in stochastic
benchmarks~\cite{danon2005comparing, lancichinetti2009community}. 
We refer to the approximation 
obtained with the Shannon's theorem as the
decodability bound, being this quantity always a lower bound of the 
exact recovery threshold. 
The derivation of the decodability bound 
is very simple, as it requires only estimating the channel capacity of a
generative network model. Further, the bound appears to be not very sensitive
to important ingredients of a generative model with planted community
structure such as degree and community size distributions.
The procedure seems therefore potentially generalizable 
to more complicated models. 
\change{Please note that the 
decodability bound is still derived under the hypothesis that network
size is infinite, thus its effectiveness in
  finite-size systems is not a priori granted. However,
numerical results  
provide compelling evidence that no algorithm has been able to 
beat the decodability bound in 
the settings of the stochastic block model generally
considered in the literature~\cite{danon2005comparing, lancichinetti2009community}.
At the same time,
we note also that the best algorithms available on the market have
performances already very close to those predicted by the decodability
bound, leaving therefore only litte room for
potential improvement.} 
To reach our main results, 
we follow a path at the interface between community detection 
and traditional coding theory~\cite{mackay2003information}.
In particular, we deploy a family of error-correcting 
algorithms for community detection that 
rely on a traditional decoding technique~\cite{gallager1962low,
  mackay1996near}. Although these methods do not provide
scalable algorithms, they do
allow for straightforward calculations showing that the
decoding algorithms are theoretically able to achieve 
almost perfect recovery at the decodability bound.

\section{Methods}

\subsection{Low-density parity-check codes}

In this section, we describe a family of linear codes that
can be used to interpret the problem of community detection in
networks from a classical information-theoretic perspective.
We stress that the code $\mathcal{P}$ that we are going to
present has been originally introduced in Ref.~\cite{abbe2016exact}.
Here, in addition to rephrasing and expanding 
its description, we show that
the code is only one member of a 
large family of equivalent codes.

Except for the final section, the entire paper
focuses on the case of two communities only. This fact allows us
to work with a binary alphabet and modulo-$2$ arithmetic.
The message that the source wants to deliver is a vector 
$\vec{\sigma}^T = (\sigma_1, \ldots, \sigma_N)$ of $N$ bits,
where $\sigma_i = 0, 1$ specifies the
membership of node $i$, and $N$ is the size of the graph. 
We refer to
the bits $\sigma$  as the information bits. 
Note that  the transformation
$\sigma_i \to (\sigma_i + 1) $ doesn't change the 
content of information. This means that
the effective number of 
information bits generated by the source 
is $N-1$. The encoder acts on the message in
a very simple way. It basically generates a network
where communities are disconnected cliques. From this perspective,
edges are purely redundant information, as the
encoder generates $N(N-1)/2$  bits for every pair $(i,j)$ of nodes to
satisfy parity-check equations of the type
\begin{equation}
\sigma_i + \sigma_j + \theta_{i,j}  = 0 \;.
\label{eq:encoder1}
\end{equation} 
The bits $\theta$ are called the parity bits of the codeword.
Note that, in any codeword, 
an edge corresponds to a parity bit $\theta = 0$, and every
non-edge to a parity bit $\theta = 1$. This choice is made
for convenience. 
 We refer to the code  described by the system of 
 Eqs.~(\ref{eq:encoder1})  as the pair code, or shortly as the code
 $\mathcal{P}$. 
The rate
of this code is
\begin{equation}
R = \frac{\log_2 (2^N/2)}{N(N-1)/2} = \frac{2}{N}
\label{eq:rate}
\end{equation}
as the total number of possible messages is $2^N/2$ (the division
by $2$ arises from symmetry), and the total number of parity bits is
$N(N-1)/2$.
The code $\mathcal{P}$ is linear, and
can be written
as a single matrix-vector equation $H \vec{\chi} = \vec{0}$, where
$\vec{\chi}^T = (\sigma_1, \ldots, \sigma_N, \theta_{1,2}, \ldots,
\theta_{N-1, N})$ is the codeword, $\vec{0}$ is a vector with the same
dimension
as $\vec{\chi}$ 
but where every single component is equal to zero, and
\begin{equation}
H = \left(
\begin{array}{c|c}
V^{T} & \mathbb{I}_{N(N-1)/2} 
\end{array}
\right) 
\label{eq:parity-check}
\end{equation}
is the parity-check matrix of the code. 
In the above expression, 
$\mathbb{I}_{q}$ is the
 identity matrix of dimensions $q \times q$, whereas 
$V$ is a rectangular matrix with $N$ rows and $N(N-1)/2$ 
columns that
can be written as composed of $N$ blocks
\[
V = \left( \begin{array}{c|c|c|c|c|c}
V_1 & V_2 & \cdots & V_i & \cdots & V_{N}
\end{array}
\right) \; .
\]
The $i$-th block is defined as
\[
V_i = \left( \begin{array}{c}
0_{ [i-1] \times [N-i]}
\\
J_{1 \times [N-i]}
\\
\mathbb{I}_{N-i}
\end{array}
\right) \; ,
\]
where $0_{q \times t}$ is a matrix with $q$ rows and 
$t$ columns whose entries are all equal to zero, and
$J_{q \times t}$ is a matrix with $q$ rows and 
$t$ columns whose entries are all equal to one. 
The parity-check matrix of the code $\mathcal{P}$ has two nice properties.
First, it appears in the so-called systematic
form. 
This means that
the actual generator matrix $G$ of the code, the one used to generate
codewords as $\vec{\sigma}^T G = \vec{\chi}^T$, can be written
as $G = \left( \mathbb{I}_{N} | V \right)$. Second, 
$H$ is a sparse matrix, as the density 
of ones is vanishing as $N$ grows. Linear codes based on 
low-density parity check matrices are usually denoted as LDPC codes,
and they are at the basis of many error-correcting 
techniques~\cite{gallager1962low, mackay1996near}.

After the message is encoded, the 
codeword $\vec{\chi}^T = (\vec{\sigma}^T, \vec{\theta}^T)$
is sent through a noisy communication channel. The effect of the channel is
twofold. First, it erases completely the information bits
$\sigma$. Second, it changes the value of some parity bits $\theta$.
 What is received at the end of the channel
is therefore a network with only partial information about
the original community structure generated by the source.
The way one can attempt to recover the content of the original 
message is finding the codeword that 
best represents, in terms of minimal
distance, the 
received word. Several decoding algorithms can 
be used in this respect. 
A naive approach is
for instance based on a spectral algorithm (see Appendix~\ref{app:1}).
In this paper, we consider 
state-of-the-art 
error-correcting 
algorithms, typically used in decoding
processes over arbitrary memoryless noisy channels~\cite{mackay2003information}.
We will come back to it in a moment. 
Meanwhile, we would like to make 
some important remarks.

Although apparently very similar,  
the interpretation presented here is different from
the one considered by Rosvall and
Bergstrom~\cite{rosvall2007information}.
The two approaches suffer from the 
detectability limit
in the stochastic block model~\cite{peixoto2013parsimonious}. 
They stand, however, for different takes of the 
community detection problem.
Their difference is analogous to the one present between
the source coding theorem and the
noisy-channel coding theorem~\cite{mackay2003information}.
In Ref.~\cite{rosvall2007information},
the authors rephrased the community
detection problem as a communication task
over a channel with limited capacity. 
The goal of their approach was to provide 
the best encoding strategy to deliver information
with such a limitation. Here instead, the focus is on the performance
of the communication task depending on the noise
of the channel. In this respect, it is very important to remark 
that, in most of practical situations,
one has no clue of the type of noise that characterizes  
the channel. In these situations, the only possibility is 
to make and test hypotheses. This is pretty much in  
the same spirit as of community identification
algorithms based on statistical inference~\cite{decelle2011inference,
karrer2011stochastic}, or,
from the perspective of coding theory,
maximum-likelihood decoders
devised for specific noisy channels. We are not seriously concerned
by the lack of knowledge about the
noise of the channel, as the main goal of the paper 
is to generate insights
on the problem of community detection in 
networks, rather than simply
deploying practical algorithms.

The fact that the code is linear has one important
feature~\cite{bierbrauer2005coding}. 
One can create equivalent linear codes by performing special
types of operations on the matrix $H$, as for example
permutation of rows and columns, multiplications
of rows by non-zero scalars, sum of rows, 
and so on.  Equivalence means that the codes
share the same set of codewords.
For instance, summing the rows
corresponding to the equations $\sigma_i + \sigma_j + \theta_{i,j}  =
0$, $\sigma_i + \sigma_k + \theta_{i,k}  = 0$, 
and $\sigma_j + \sigma_k + \theta_{j,k}  = 0$, 
one obtains the equation  
\begin{equation}
\theta_{i,j} + \theta_{i,k}
+ \theta_{j,k} =0 \;.
\label{eq:code_t}
\end{equation} 
This equation involves only parity bits and
not information bits. One can actually apply the same operation to 
all triplets of nodes to obtain an equivalent system of equations
consisting in sums of triplets of parity bits only. 
We refer to this as the triplet code, or shortly as the code
$\mathcal{T}$. The equivalence between
$\mathcal{P}$ and $\mathcal{T}$ is particularly
simple. If we find a codeword
containing only parity bits for $\mathcal{T}$, we
are able to trivially deduce the information
bits of the corresponding codeword of $\mathcal{P}$.
This can be done by simply fixing the value of
one information bit $\sigma_{i^*} = 0, 1$, and use
Eq.~(\ref{eq:encoder1})
to iteratively retrieve the values of the 
other information bits such that they satisfy their respective
parity-check equation.
This fact tells us
that looking at higher-order local structures 
doesn't provide any significant benefit
to the decoding process, aka 
the identification of the communities in the network. 
We will see, however, that working 
with the code $\mathcal{T}$
is useful to characterize performances of decoders.

\subsection{Gallager decoders}
A convenient way to graphically represent a LDPC code is to use a bipartite
network called Tanner graph. The graph is constructed
from the parity-check matrix $H$ of the code. Every row of $H$
identifies a check node, and every column of $H$ identifies 
a variable node. A variable node $v$ is connected
to a check node $c$ if the entry $H_{c,v} =1$; if $H_{c,v} = 0$, $v$ and $c$
are instead disconnected. 
Tanner graphs are particularly 
useful in the description of a probabilistic 
decoding strategy introduced by Gallager in the
1960s~\cite{gallager1962low}. 
We provide in the Appendices~\ref{app:pair} and~\ref{app:triplet}
 a detailed description of the algorithm
as implemented for the specific cases of the 
$\mathcal{P}$ and $\mathcal{T}$ codes, respectively.
Here, we just
describe the spirit of the approach, and report the simplified
equations for the code $\mathcal{P}$.
The technique consists in a series of messages
exchanged by check nodes and variable nodes connected
in the Tanner graph. A variable node $v$
sends to a connected check node $c$ a message  
$m_{v \to c}$ representing the probability of the bit value
that $v$ should assume according to the other check 
nodes $c' \neq c$ connected to it.
In turn, a check node $c$ replies to a variable node $v$ with a
message $n_{c \to v}$ consisting in the probability
of the bit value that the other variable nodes $v' \neq v$ attached to $c$
would like to see from $v$. The algorithm is initialized from suitable
initial conditions, i.e., our beliefs on the variable nodes, 
and run until convergence or up to a 
maximum number of iterations.  The approach is exact in acyclic
Tanner graphs, and thus particularly
effective in the context of LDPC codes. Cycles in the Tanner
graph deteriorates the performance of the algorithm, as they 
introduce dependencies among messages that are actually 
neglected in the algorithm by Gallager. Most of the information
theory research 
in this context is indeed centered on the construction of 
LDPC matrices (not
necessarily equivalent)  with small number of short loops, avoiding
in particular loops of length four.

Going back to our specific problem, the matrix $H$ defined in
Eq.~(\ref{eq:parity-check}) generates a Tanner graph with girth equal
to six (the girth is the length of the shortest loop in the
graph). Such a property cannot be changed by creating equivalent
parity-check matrices. In principle, one can apply the Gallager
algorithm to any of the Tanner graphs generated starting 
from equivalent parity-check matrices, leading therefore to a class
of algorithms. Typically, the more irregular, in terms of degree
for check and variable nodes, the Tanner 
graph is, the lower is the total
number of iterations required for the eventual convergence. On the other
hand, increasing the degree heterogeneity of the Tanner
graph increases also the
complexity of the algorithm, whereas the solution 
obtained by the algorithm
remains basically the same. If one applies the Gallager 
algorithm to the Tanner graph generated
from the systematic parity-check matrix 
of Eq.~(\ref{eq:parity-check}), it is possible
to simplify the implementation
of the algorithm (Appendix~\ref{app:pair}), and obtain 
the following system
of equations that relates variables at iteration $t$ of the
algorithm to the values of the same variables at stage $t-1$
of the algorithm:
\begin{equation}
\zeta^{(t)}_{i \to j} = \left\{
\begin{array}{ll}
\ell_i &  \textrm{ , for } t =0
\\
\ell_i  + \sum_{k \neq j \neq i}    \log \, \frac{1 + \, \tanh{(1/2 \, 
    \zeta^{(t-1)} _{k \to i}  )}  \, \tanh{(1/2 \, 
    \ell_{i,k}  )}   }  {1 - \, \tanh{(1/2 \, 
    \zeta^{(t-1)} _{k \to i}  )}  \, \tanh{(1/2 \, 
    \ell_{i,k}  )}  }   &  \textrm{ , for } t \geq 1 
\end{array}
\right. \; ,
\label{eq:mp}
\end{equation}
where the logarithm is taken in the natural basis, and $\tanh(\cdot)$ is the
hyperbolic tangent function. 
$\ell_i = \log \frac{P(\sigma_i=0|s_i)}{P(\sigma_i=1|s_i)}$
is the {\it a priori} log-likelihood ratio (LLR) of the information bit
$\sigma_i$, based on the information received
$s_i$.  It essentially stands for the logarithm of the ratio between the probabilities
that  $\sigma_i$ is zero and $\sigma_i$ is one, given our observation $s_i$ at
the end of the noisy channel.  $\ell_{i,j} = \log
\frac{P(\theta_{i,j}=0|A_{i,j})}{P(\theta_{i,j}=1|A_{i,j})}$
is exactly the same quantity but for the parity bit $\theta_{i,j}$,
and $A_{i,j}$ is the element $(i,j)$ of the adjacency matrix of the
graph, i.e., one plus the $(i,j)$-th parity bit received at the end of the channel.
The messages $\zeta^{(t)}_{i \to j}$  are also
LLRs. They are defined for every pairs of nodes $i$
and $j$, not just those actually connected. 
At every iteration $t$, the best estimate
of the LLRs for the information and
parity bits are given respectively by
\begin{equation}
\hat{\ell}^{(t)}_{i} = 
\ell_i  + \sum_{k \neq i}    \log \, \frac{1 + \, \tanh{(1/2 \, 
    \zeta^{(t-1)} _{k \to i}  )}  \, \tanh{(1/2 \, 
    \ell_{i,k}  )}   }  {1 - \, \tanh{(1/2 \, 
    \zeta^{(t-1)} _{k \to i}  )}  \, \tanh{(1/2 \, 
    \ell_{i,k}  )}  } \; ,
\label{eq:best1}
\end{equation}
and 
\begin{equation}
\hat{\ell}_{i,j}^{(t)} = \ell_{i,j} +   \log \, \frac{1 + \, \tanh{(1/2 \, 
    \zeta^{(t-1)} _{j \to i}  )}  \, \tanh{(1/2 \, 
    \zeta^{(t-1)} _{i \to j}   )}   }  {1 - \, \tanh{(1/2 \, 
    \zeta^{(t-1)} _{j \to i}  )}  \, \tanh{(1/2 \, 
    \zeta^{(t-1)} _{i \to j}   )} } \; .
\label{eq:best2}
\end{equation}
A hard-decision
process is made at every stage $t$: the best estimates of the 
information bit $i$ is $\hat{\sigma}_i = 0$ if
$\hat{\ell}_{i,j}^{(t)}>0$,
and $\hat{\sigma}_i = 1$, otherwise. Similar rules hold for the
determination of the
best estimate of the 
parity bits $\hat{\theta}$. These values are used to test
convergence 
of the algorithm by simply checking whether the best estimates of the
parity
and information bits are such that all Eqs.~(\ref{eq:encoder1})
are satisfied. In such a case, the algorithm stops. Otherwise, the algorithm is run up to a maximum number
of iterations.
The {\it a priori} LLRs $\ell_i$ and 
$\ell_{i,j}$  appearing in Eqs.~(\ref{eq:mp}), 
~(\ref{eq:best1}), and~(\ref{eq:best2})
play a fundamental role 
as they determine the fixed point which the algorithm will converge to.
These parameters represent our prior belief about
the message sent from the source on the basis of the corrupted
word we received on the other side
of the channel. If no other information is available,
we can simply set $\ell_i = 0$ for all $i$, except for
a single node $i^*$ for which $\ell_{i^*} = \pm \infty$.
The latter condition is given by the fact that original
message is perfectly symmetric, so that we can impose
that a given information bit is certainly $0$ (i.e., $\ell_{i^*} = +
\infty$) or $1$ (i.e., $\ell_{i^*} = -
\infty$) without providing aid to our
decoder. The values of the 
LLRs $\ell_{i,j}$
are instead functions of the features of the
noisy channel, and the observed value of the
element of the adjacency matrix $A_{i,j}$. 
Without any prior knowledge (or hypotheses)
on how the parity bits were actually
altered by the channel, the approach is thus
not applicable. 

The iterative algorithm just introduced resembles
the one considered by Decelle {\it et al.}~\cite{decelle2011inference,
decelle2011asymptotic}. This is a natural consequence of the fact that both 
algorithms are trying to solve
the same type of problem. There are, however, some
differences. The most important one is conceptual,
as ours is a straight adaptation of a well-established decoding technique
to a specific decoding tasks. In this respect, the algorithm
maintains a general character. For instance, the algorithm explicitly
adapts to any noisy channel by simply choosing appropriately the
values of LLRs $\ell$.
Also within the same noisy channel,
the algorithm written for the code $\mathcal{P}$ 
is just one of
the potentially many algorithms that
can be generated starting from equivalent parity-check
matrices. 
Further, our algorithm includes an error-correcting component 
for the parity bit values [Eq.~(\ref{eq:best2})].
Finally, the performance of the family of algorithms 
can be studied
with a standard technique of coding theory named density evolution~\cite{gallager1962low,
  richardson2001capacity}, as we are going to
illustrate below.
We should remark, however, that the algorithm has the 
practical disadvantage of working with a number of equations that scales
quadratically with the system size, rather than
linearly as the algorithm by Decelle {\it et al.} This is a
consequence
of the general nature of the algorithm, being not devised to perform 
the specific decoding task considered here.
In this respect, we stress that other efficient 
and effective coding-theoretical algorithms specifically devised
for the stochastic block model
are available on the market~\cite{ mossel2013proof, abbe2016achieving}.


\begin{figure}[!htb]
\begin{center}
\includegraphics[width=0.48\textwidth]{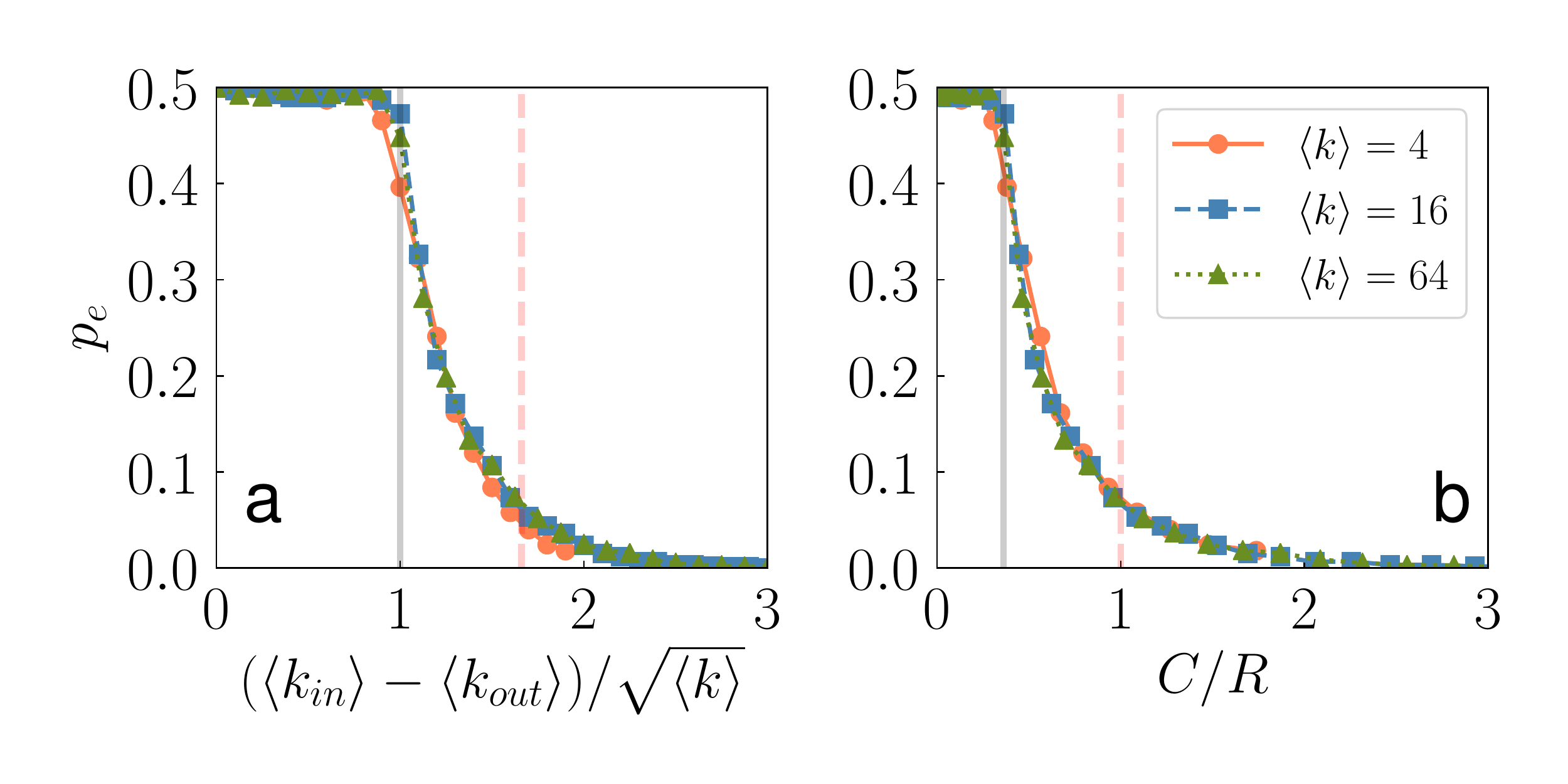}
\end{center}
\caption{(Color online)
Performance of the Gallager algorithm in the stochastic block model
with two communities. (a) The plot shows the probability of error $p_e$ for the 
information bits, i.e. the fraction of
information bits that are not correctly
decoded by the algorithm. We consider block models with 
$N = 10,000$ divided in two equally sized groups
with $n = N/2$ nodes. Each curve represents the results
obtained for fixed 
average degree $\langle k \rangle = (n-1) p_{in}
+ n p_{out}$. The probability of error is computed
as a function of difference between the average
internal degree $\langle k_{in} \rangle = (n-1) p_{in}$
and the average external degree 
$\langle k_{out} \rangle = n p_{out}$. As the figure shows, the
decoder is not able to recover any information in the regime
$\langle k_{in} \rangle - \langle k_{out} \rangle \leq \sqrt{\langle k
  \rangle}$ (gray full line). The probability of error is larger than zero
at the decodability bound (red dashed line).
(b) Same data as in panel a, but the probability of error is plotted
against the ratio $C/R$ between channel capacity and 
rate of the code.
}
\label{fig:2}
\end{figure}

\section{Results}
\subsection{The stochastic block model and the 
detectability threshold}

In terms of performance, our algorithm behaves similarly to
the one by Decelle {\it et al.}  This fact is visible in
Figure~\ref{fig:2}a, where we consider
the application of the algorithm to the stochastic block model,
finding once more the existence of the so-called detectability
threshold~\cite{decelle2011inference, nadakuditi2012graph}.
The information-theoretic sufficient and necessary condition
of the detectability threshold has been proven in 
Refs.~\cite{mossel2013proof, abbe2016exact, abbe2015community, 
abbe2017community}.
In its simplest variant (the one considered here),
the stochastic block model serves to
generate networks with planted community structure, where
$N$ nodes are divided in two groups of size 
$n$ and $N-n$, respectively.
Nodes belonging to the same 
group are connected with probability $p_{in}$, while pairs of nodes
belonging to different groups are connected with probability
$p_{out}$. Using this knowledge of the channel, we can easily estimate
the value of the LLRs $\ell_{i,j}$ required by the Gallager algorithm
(Appendix~\ref{app:llr}).  The detectability threshold is generally studied
for equally sized groups, so that $n=N/2$. 
One defines the average internal degree as $\langle k_{in} \rangle = (n-1)
p_{in}$, 
the average external degree as $\langle
k_{out} \rangle = n p_{out}$, and the average degree as $\langle
  k\rangle = \langle k_{in} \rangle +\langle k_{out} \rangle$.
If the difference $\langle k_{in} \rangle -\langle k_{out} \rangle$
is smaller than $\sqrt{\langle k \rangle}$, the algorithm
is not able
to detect any group. Groups start to be partially 
decoded only 
when $\langle k_{in} \rangle -\langle k_{out} \rangle > \sqrt{\langle
  k \rangle}$.

\subsection{The capacity of the noisy channel associated 
with the stochastic block model}

\begin{figure*}[!htb]
\begin{center}
\includegraphics[width=0.75\textwidth]{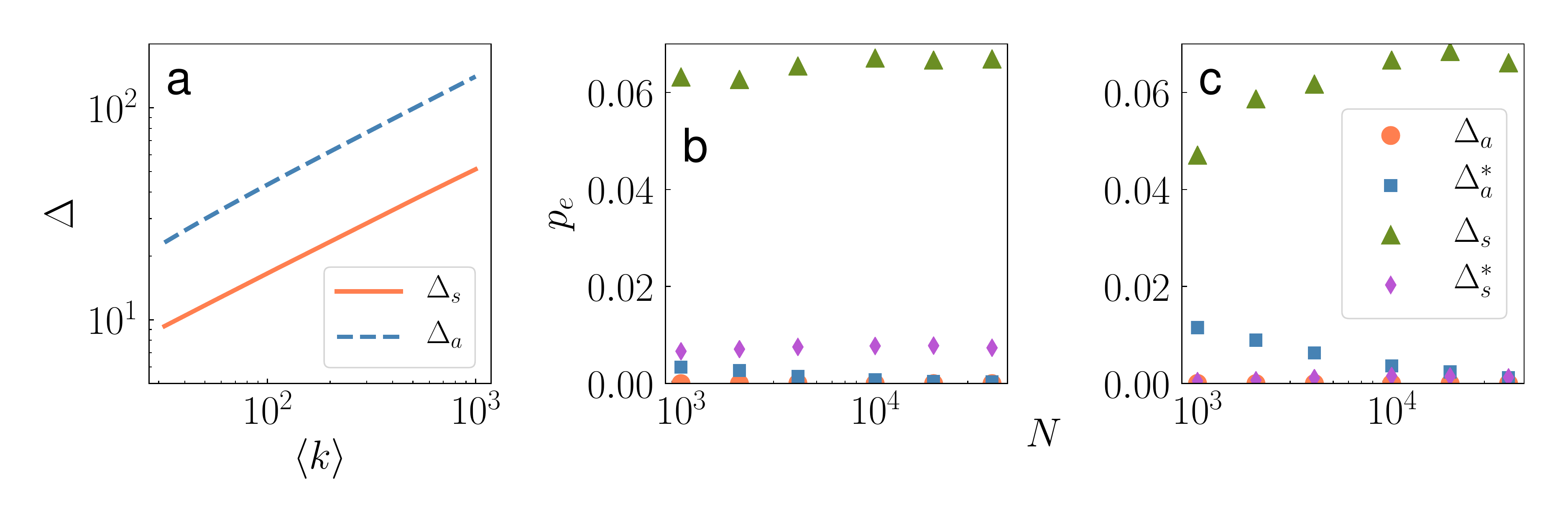}
\end{center}
\caption{(Color online) Performance of the Gallager decoder at threshold. (a)
  Comparison between the exact recovery threshold ($\Delta_a$, dashed blue
  line) and the decodability bound ($\Delta_s$, orange full line)
for a stochastic block model with $N=10,000$ nodes as a function of
the average degree $\langle k \rangle$. (b) Probability of bit error $p_e$ as
a function of the network size $N$. We fix the value of the average
degree $\langle k \rangle = 32$, and apply the Gallager decoder 
to $10$ instances of the model. Results stand for average values 
of $p_e$ over those realizations. Different
colors and shapes correspond to different choices of $\langle k_{in}
\rangle$. Orange circles are obtained at the exact recovery
threshold; blue squares correspond to regimes of slightly larger
amount of noise than the one corresponding to the 
exact recovery threshold , that is $\langle k_{in}
\rangle$ is 0.9 times of the threshold value; green triangles
are obtained at the decodability bound; purple diamonds are
obtained slightly above the decodability bound
by setting $\langle k_{in}
\rangle$ is 1.1 times of its decodability 
bound value. (c) Same as in panel b,
but for $\langle k \rangle = 128$.}
\label{fig:3}
\end{figure*}

The detectability threshold is one side of the medal. It
tells us what is the minimum level of disorder that the
channel should introduce to disrupt completely our ability to decode
the original signal. One may, however,
be interested in the behavior at regimes of lower
noise, specifically to the value of maximum noise that can be tolerated
to achieve perfect decoding, i.e., retrieve the original
signal with no mistakes. This is often the case considered in the
literature about performance of 
community detection algorithms~\cite{lancichinetti2008benchmark,
  lancichinetti2009community}. It is also the typical case
contemplated in information theory
for reliable communication~\cite{mackay2003information}.
Recent literature has shown that exact recovery in the stochastic
block model is still subjected to a threshold phenomenon~\cite{abbe2016exact, abbe2017community}. 
The value of  the threshold for the stochastic 
block model with two communities is given by
$\langle k_{in} \rangle   - \langle k_{out} \rangle   = \log N \, \sqrt{ \frac{2 \langle
    k \rangle}{\log N} - 1  } $
 (see Appendix~\ref{app:abbe} for the rephrasing of the original results of
  Refs.~\cite{abbe2016exact, abbe2017community} 
according to the notation used in the current
paper). The above condition is valid in the limit of infinitely-large
stochastic block models with two equally-sized communities 
decoded using a maximum likelihood decoder. The rationale 
behind the existence of such a finite-threshold
effect is analogous to the one that describes 
the connectivity of the
Erd\H{o}s-R\'{e}nyi model~\cite{erdos1960evolution}. 
For instance, the logarithmic dependence of the threshold from the
system size arises from the requirement of having
no nodes with degree equal to zero, as those nodes 
cannot be correctly classified. 

The threshold value is exact for infinitely large
systems. However in numerical validations of community detection
algorithms~\cite{danon2005comparing,
  lancichinetti2009community}, system sizes are not very large.
As we are going to show, the exact recovery threshold determines
a too restrictive condition that doesn't
provide an accurate estimate of the regime of exact
recovery reached by the best algorithms. 
Such a condition must be relaxed to obtain more reliable 
predictions in finite-size systems.
Here, we propose a simple way to do it.
We compute a lower bound on the true value of the threshold
using the Shannon's noisy-channel coding theorem~\cite{shannon}. 
We refer to it as the decodability bound. 
The value of the bound differs from the one 
of the exact recovery threshold for a simple reason, 
already well emphasized
in Ref.~\cite{abbe2015community}:  
Shannon's theorem is not
the mathematically correct way to study exact recovery
in community detection.
The theorem applies in fact to
the case where the channel properties are independent
of the choice of the code. In the noisy-channel interpretation
of community detection instead, the code is given, and there
is no way of playing with it without necessarily changing
the features of the channel. Given the lack of flexibility, the
decodability bound is 
necessarily a lower bound of the true threshold. This tells us that
exact recovery is impossible if the noise level is higher than what predicted 
by Shannon theorem. On the other hand, having a lower amount of noise 
than the one established by the bound doesn't provide a
sufficient condition for perfect recovery. 
Both the exact recovery threshold and the decodability bound scale
with the square root of the average 
degree of the graph (see Fig.~\ref{fig:capacity_sm1}).
The difference between them grows logarithmically with the system size
(see Fig.~\ref{fig:capacity_sm1}, and Fig.~\ref{fig:3}a). 
As Shannon's theorem is still derived in the limit of infinitely
large systems, the decodability bound is potentially subjected to the
same limitations as those of the exact recovery threshold. 
However, the different scaling
with the system size of the decodability bound makes it 
a meaningful indicator for characterization of performances
of community detection
algorithms~\cite{danon2005comparing,
  lancichinetti2009community}.
Another nice feature of the decodability bound is that it can 
be computed in a rather simple way by
knowing just the properties of the stochastic network model
without relying on any specific decoding protocol.
In the following, we provide concrete support to these
  statements. For simplicity, we start describing how
the decodability bound is computed in the stochastic block model
with two communities. We will then proceed with calculations valid
for other models, and  numerical tests of the predictive power of
the decodability bound.

In the interpretation of the community detection problem
as a communication process, the
stochastic block model is equivalent to
an asymmetric binary channel~\cite{neri2008gallager}.
We can compute the capacity of the channel as (see Appendix~\ref{app:capacity})
\begin{equation}
C  = H_2 [ \alpha^* p_{in} + (1- \alpha^*) p_{out} ] - \alpha^* H_2(p_{in}) - (1-\alpha^*)
H_2(p_{out}) 
\label{eq:cap}
\; .
\end{equation}
$H_2(x) = - x \log_2 x - (1 - x) \log_2 (1-x)$ is the binary
entropy function. $\alpha^*$ is a function
of $p_{in}$ and $p_{out}$, and represents the value
of the proportion of parity bits $\theta=0$ in the transmitted
codeword that maximizes the mutual information between
transmitted and received words. 
Knowing the capacity of the channel $C$ and the 
rate $R$ of the code
[Eq.~(\ref{eq:rate})], the decodability bound is
determined by the condition $C/R = 1$.
As the results of Figure~\ref{fig:2}b show, the
ratio $C/R$ provides a natural scale
to monitor the performance of the algorithm
as a function of the channel noise, at the same footing as
$(\langle k_{in} \rangle -\langle k_{out} \rangle) /  \sqrt{\langle k
  \rangle}$ does.
The two quantities are effectively related by the law 
$C/R \sim (\langle k_{in} \rangle -\langle k_{out} \rangle)^2 /
\langle k \rangle$
(see Fig.~\ref{fig:capacity_sm1}).

The results of Figures~\ref{fig:3}b and ~\ref{fig:3}c 
highlight one of the potential reasons of why the
decodability bound turns out to be more informative 
than the exact recovery threshold in finite-size systems.
The figures show how the Gallager algorithm  
performs at different levels of noise 
as the system size increases.
Slightly above the decodability bound but below 
the exact recovery threshold, the probability of information-bit error $p_e$
is not exactly equal to zero.  However, $p_e$ is so small 
to become unnoticeable in numerical simulations. Further, 
the entity of the error becomes even smaller as 
the average degree
increases.

\begin{figure*}[!htb]
\begin{center}
\includegraphics[width=0.85\textwidth]{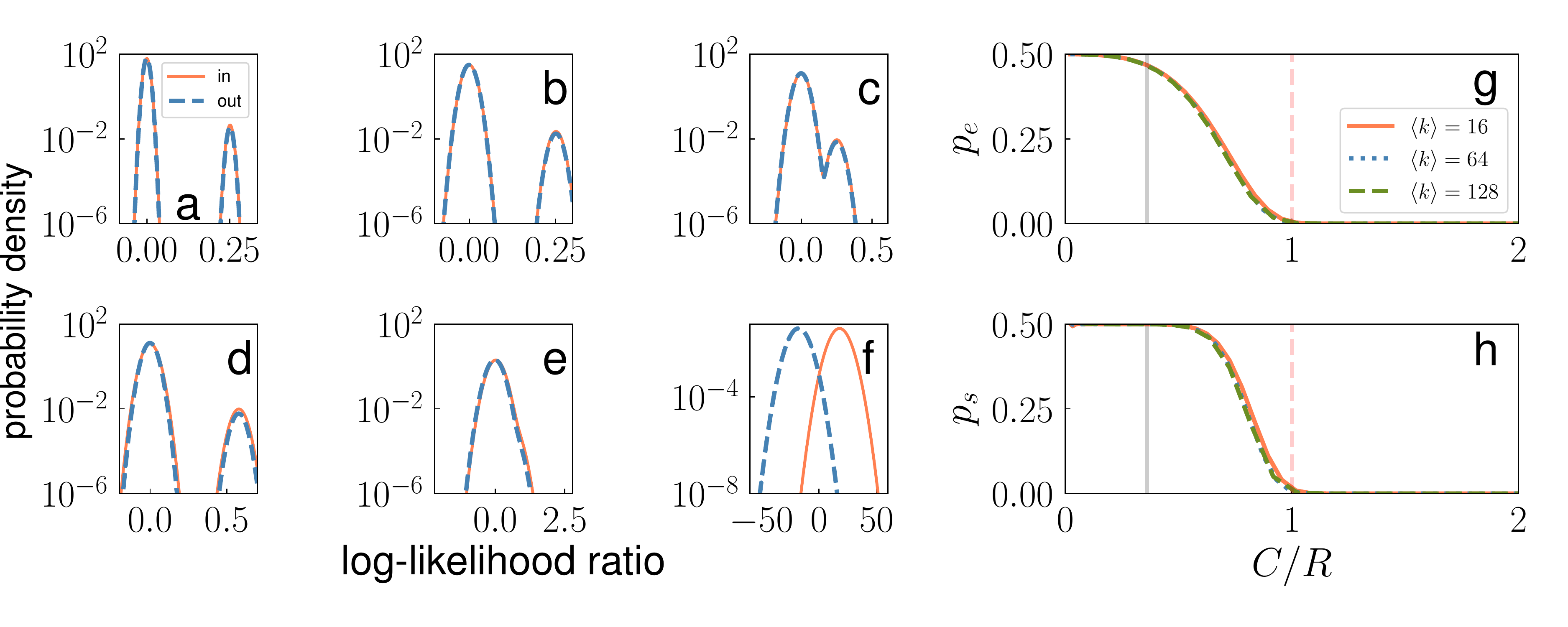}
\end{center}
\caption{(Color online) Achieving the capacity of the stochastic block
  model. 
We consider a stochastic block model with $N=10^5$, and
  groups of equal size $n=N/2$.
(a) Probability density of the log-likelihood ratios (LLRs) after one
iteration of the Gallager algorithm. The full orange line indicates
the
distribution for parity bits corresponding to internal pairs of nodes.
The dashed blue line  indicates the distribution for parity bits
corresponding to external 
pairs of nodes. Here, we set $\langle k _{in} \rangle = (n-1) p_{in} =
36$, and $\langle k _{out} \rangle = n p_{out} =
28$. The parameter values
correspond to the regime where communities are not
detectable. (b and c) Same as in panel a, but for the second and third iterations
of the algorithm. (d, e, and f) Same as in panels a, b and c, but for 
$\langle k _{in} \rangle = (n-1) p_{in} =
41$, and $\langle k _{out} \rangle = n p_{out} =
23$. In this regime communities are detectable,
in the sense that good algorithms should be
able to identify a non-vanishing portion of the nodes
belonging to the two clusters. (g) Performance of the iterative algorithm after three
iterations. We consider three different values of the average degree
$\langle k \rangle$. The description of the figure is similar to the
one of Fig.~\ref{fig:2}b with the difference that the probability of
error $p_e$ refers to parity bits only.
(h) Probability $p_s$ that a randomly chosen parity-check equation of
the code $\mathcal{T}$ is not satisfied
as a function of the ratio $C/R$. We consider the same networks as
those of panel g.
}
\label{fig:4}
\end{figure*}

\subsection{Capacity-achieving codes}
An issue that still remains open is understanding the performance of
the LDPC codes we introduced earlier in the paper from a more formal
point of view. In particular, we would like to better
characterize their performances around the decodability bound.
To address the issue, we rely 
on a popular technique, called density evolution,
generally used to study the performances of LDPC decoders based on the
Gallager algorithm. 
The advantage of the approach is that it allows us to 
study the performance of the algorithm without the need to run any simulation.
The technique was introduced by Gallager
himself in the analysis of the binary symmetric 
channel~\cite{gallager1962low}, and later
generalized to arbitrary channels by 
Richardson and Urbanke~\cite{richardson2001capacity}. 
The technique
consists in assuming the LLR of the variable nodes as independent,
and study the evolution of their probability density
during the first $t$ stages of the
algorithm, with $t$ smaller or equal than half of the girth of the
underlying Tanner graph, i.e., till the assumption of independence
among variables is justified. Estimating how LLR densities evolve
requires repeated convolutions of distributions.
The mathematical treatment of the density evolution 
for Gallager algorithm for the code $\mathcal{P}$ is quite
involved. However, for the equivalent code $\mathcal{T}$, it is
greatly simplified (see Appendix~\ref{app:density}). In the limit of sufficiently large 
$N$, the distributions
$P^{(t)}_{in} (\hat{\ell})$ and $P^{(t)}_{out} (\hat{\ell})$,
respectively valid for the LLR of parity bits corresponding to 
pairs of nodes in the same group (internal pairs) and different groups
(external pairs), are computed
iteratively as convolutions of normal and delta distributions.
These computations can be efficiently performed via
numerical integration whose computational cost is virtually 
independent of the system size. From the LLR densities, 
we can further estimate (i) the probability of error on the best 
estimates of the parity bits, and (ii) the probability of
error for the parity-check equations of the code $\mathcal{T}$ [Eq.~(\ref{eq:code_t})]. 
These quantities serve to judge the overall
performance of the algorithm. 
Everything can be carried out till $t=3$ iterations,
as the girth of the Tanner graph of the code equal six. 
This is a low number, yet sufficient
to capture the general behavior of the algorithm.

Results of the density evolution 
analysis are reported in Figure~\ref{fig:4}. 
First, we grasp why the detectability threshold 
emerges in the Gallager decoder.
For $\langle k_{in} \rangle
-\langle k_{out} \rangle <   \sqrt{\langle k
  \rangle}$, $P^{(t)}_{in} (\hat{\ell})$ and $P^{(t)}_{out}
(\hat{\ell})$ are essentially identical, leading to the impossibility
to properly disentangle internal from external 
parity bits.
For $\langle k_{in} \rangle
-\langle k_{out} \rangle >   \sqrt{\langle k
  \rangle}$ instead, the two distributions progressively separate
one from the other, leading to partial (or even complete)
recovery of the correct value of the parity bits. We note that
decoding correctly a portion of the parity bits does 
not necessarily correspond to
the correct recovery of a portion of the information bits (Fig.~\ref{fig:4}g). 
If some of the parity-check equations of the $\mathcal{T}$ code are
violated (Fig.~\ref{fig:4}h), then  some parity-check equations of the code
$\mathcal{C}$ are violated too. The relation between the two codes
in terms of syndromes is not trivial. Hence, one cannot conclude that
a probability of
parity-bit error smaller than $0.5$ corresponds to a
probability of information-bit error smaller than $0.5$.
On the other hand, $\mathcal{P}$ and  $\mathcal{T}$
share the same codewords, thus, if the Gallager algorithm
on the $\mathcal{T}$ code converged finding a codeword, then 
convergence to the same codeword 
is guaranteed also for the code  $\mathcal{P}$. In particular,
if the codeword for $\mathcal{T}$ is the one that perfectly disentangles parity bits
corresponding to internal and external pairs, then the 
corresponding codeword for $\mathcal{C}$ is the one that
recovers perfectly the true values of
information bits $\sigma$. As figure~\ref{fig:4}h shows, this situation
happens approximately at the decodability bound, where both
the probabilities of error for parity bits and parity-check
equations are very close to zero. As a consequence, the algorithm 
is able to achieve performance very close to the channel capacity.

\subsection{Stochastic block models with more than two groups}

\begin{figure}[!htb]
\begin{center}
\includegraphics[width=0.48\textwidth]{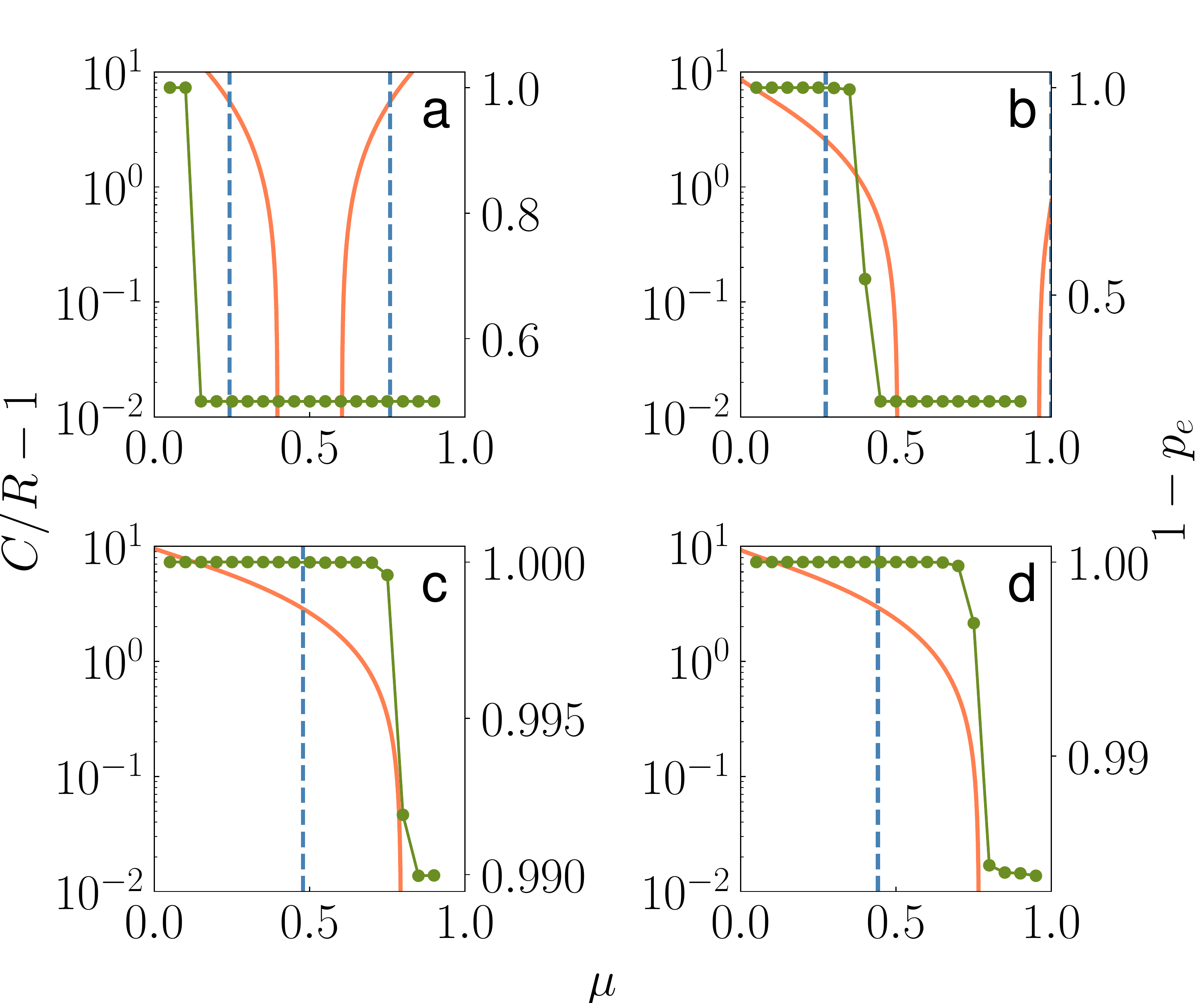}
\end{center}
\caption{(Color online)  Decodability bounds for the stochastic block model with
  multiple groups. We consider synthetic networks 
where $N$ nodes are divided into $Q$ groups of equal size. Pairs of
nodes within the same group are connected with probability $p_{in}$,
and pairs of nodes belonging to different groups are connected with
probability $p_{out}$. The average value of the internal degree of a
node is $\langle k_{in} \rangle = p_{in} (N/Q -1) $, whereas the
average
value of the external degree is $\langle k_{out} \rangle = p_{out} N
(Q -1) /Q $. The probabilities $p_{in}$ and $p_{out}$ are subjected to the constraint
$\langle k \rangle = \langle k_{in} \rangle + \langle k_{out}
\rangle$. 
As the decodability
bound is determined by the condition $C/R =1$, with 
$C/R$ ratio between the capacity of the channel and
the rate of the code, we plot $C/R -1$
 as a function of the mixing parameter $\mu =
\langle k_{out} \rangle / \langle k \rangle$. The latter quantity is
used in place of the difference $\langle k_{in} \rangle - \langle k_{out} \rangle$
to make the results easily interpretable in the comparison with the
performances of community detection algorithms 
on the same model~\cite{lancichinetti2009community}. 
The decodability bound 
is determined by the $\mu$ value where the orange
full line drops down to values smaller than zero. 
As a term of comparison we plot, as vertical dashed blue
lines, the value of exact recovery threshold~\cite{abbe2015community}
(see Appendix~\ref{app:abbe} for details on how the threshold is computed).
As a paradigmatic example of a good community detection
algorithm, we used Infomap~\cite{rosvall2008maps}, i.e., the
top-performing algorithm according to the analysis of
Ref.~\cite{lancichinetti2009community}. 
For every $\mu$ value, performance is measured in terms of $1
- p_e$, where $p_e$ is the probability of parity-bit error 
(green squares). Exact recovery correspond to $p_e=0$.
The results presented in the various panels 
refer to the average value of
$p_e$ computed over at least  $10$ independent 
realizations
of the synthetic network model.
(a) As in Fig.~\ref{fig:2}, we set $N=10,000$, 
$Q=2$ and $\langle k
\rangle = 64$. (b) Same as in panel a, 
but for $N=128$, $Q=4$ and
$\langle k \rangle =16$.
(c) Same as in panels a and b but for $N=5,000$, $Q=100$ and $\langle k
\rangle = 20$. (d) Same as in panels a, b, and c, 
but for $N=5,000$, $Q=50$ and $\langle k
\rangle = 20$.
For the computation of the exact recover threshold and
the decodability bound, parameter values in panels c and d are chosen such that
they are comparable with those used in Figures~1 and ~2 
of Ref.~\cite{lancichinetti2009community}. In panels c and d,
  Infomap is run on the LFR benchmark. Parameters of the model are
  chosen identical to those considered
  in Ref.~\cite{lancichinetti2009community}.
}
\label{fig:5}
\end{figure}

So far, we considered the simplest scenario of stochastic block models
composed of two groups only. This is a rather special
case, as the problem  of identifying the memberships of the nodes can be
mapped into a linear decoding task. Writing
linear codes that apply to stochastic block models with more than two groups 
seems challenging. However, we can 
still provide insights to the problem of community 
detection by simply studying the channel
characteristics, and relying on the Shannon's noisy-channel coding theorem to provide
a lower bound for the maximal amount of noise
admitted for exact recovery. 
From the graphical point of view,
the situation of multiple groups is identical to the one of two
groups (Fig.~\ref{fig:1}): we can immagine that the 
encoder generates a network of
disconnected cliques, where every clique corresponds
to a planted community. 
The effect of the channel is also the same as for the case of two
groups: 
it erases completely the information bits, and flips the values of
some of the parity  bits according to some stochastic rule.  If the
number of groups is $Q$, the rate
of the code is given by
\begin{equation}
R = \frac{\log_2 (Q^{N}/Q)}{ N(N-1)/2} = \frac{2 \log_2 Q}{N} \; ,
\label{eq:rate_multi}
\end{equation}
i.e., the generalization of Eq.~(\ref{eq:rate}) to the 
case of $Q$ communities. We assume that the channel
is still a binary asymmetric one, where parity bits $\theta=0$ are
flipped
with probability $1 - p_{in}$, and parity bits $\theta=1$ with
probability $p_{out}$. This scenario includes naturally the case of the 
Girvan-Newman benchmark graphs~\cite{girvan2002community}. 
Under these circumstances, we can extend all calculations
valid for $Q=2$ to arbitrary values of $Q$, arriving to the same
expression for the capacity of the channel [Eq.~\ref{eq:cap}].
The only formal difference is in the value of $\alpha^*$ (see Fig.~\ref{fig:capacity_sm2}).
The computation of this quantity requires to take
derivatives with respect  $Q-1$ variables. However, the profile of 
mutual information is pretty much flat, reaching a maximum for
a big number of different configurations. This fact allows us to
assume that the maximum of the mutual information is also reached for
equally sized groups, so that we can use
\begin{equation}
\alpha^* = \frac{N/Q-1}{N-1} \; .
\label{eq:alpha_multi}
\end{equation}

In Figure~\ref{fig:5}, we establish the value of the decodability
bound for various models of interest in the literature about
performance of algorithms in the detection of communities in 
syntethic graphs~\cite{danon2005comparing, lancichinetti2009community}.
In panel B, we consider the case of the Girvan-Newman (GN)
benchmark graphs~\cite{girvan2002community}. According to the Shannon's
theorem, decoding exactly the community memberships is impossible as
long as $C/R < 1$, with $C$ computed using Eqs.~(\ref{eq:cap})
and~(\ref{eq:alpha_multi}), and $R$ defined in
Eq.~(\ref{eq:rate_multi}). We estimate the bound in terms of the
mixing parameter $\mu =
\langle k_{out} \rangle / \langle k \rangle$
to make the results directly interpretable in terms of the numerical
tests about performances of community detection algorithms 
on the same model [see Fig.~1 of
Ref.~\cite{lancichinetti2009community}].
The results of Figure~\ref{fig:5} are particularly illuminating in this
regard. There are two regimes for which communities are in principle
perfectly decodable: (i) a sufficiently assortative regime, where
nodes have a number of internal  edges that is sufficiently
larger than the number of external connections, and  (ii) a strongly
disassortative regime, where external connections
greatly outnumber internal connections. The bound in the
assortative regime is located at $\mu \gtrapprox 0.5$.
 Indeed, the best performing algorithms in the analysis of
Ref.~\cite{lancichinetti2009community} achieve almost perfect 
recovery till $\mu = 0.4$, while their performance
drops down before reaching the point $\mu = 0.5$.
The exact recovery threshold provides instead 
much more restrictive conditions~\cite{abbe2015community}.
According to it, 
the maximum amount of noise tolerable by the channel corresponds to
$\mu \simeq 0.25$. However, most of the community detection algorithms
are able to perfectly recover the planted community structure in the
GN model well above $\mu \simeq 0.25$. 
To provide clear evidence of this fact, we replicated the
  results of Ref.~\cite{lancichinetti2009community} for
Infomap~\cite{rosvall2008maps}, i.e., one of the
top-performing algorithms.
We extended the analysis also to the 
Lancichinetti-Fortunato-Radicchi (LFR) benchmark
graphs~\cite{lancichinetti2008benchmark}, although this 
model is not exactly described by a stochastic block model (or a
binary asymmetric channel
from the perspective of coding theory).
Yet, we are able to recover approximate estimates about the regime
of perfect performance of algorithms that well describe their
behavior. In Figure~\ref{fig:5}c and~\ref{fig:5}d for instance, we estimate the
decodability regime for a stochastic block model with 
parameters that make the model comparable with the LFR benchmarks
of type $S$ and $B$, respectively, as defined in
Ref.~\cite{lancichinetti2009community}.
First, we recover threshold values that are just slightly greater
than those measured for the best performing algorithms. The
decodability bound is $\mu \gtrapprox 0.75$, point at which all
algorithms fail to correctly recover the planted partition.
Second, as the threshold value for benchmarks of type $B$ is smaller 
than the one found for benchmarks of type $S$, we are able to explain
why such a slight difference in performance is also visible in 
practical algorithms [see also Fig.~2 of
Ref.~\cite{lancichinetti2009community}]. 
A third and final deduction from the plots
is the disappearance of the disassortative regime of decodability
visible instead in the GN benchmark graph.
Still for the LFR model,  the exact recovery threshold seems to 
not well represent the regime of perfect performance of the best algorithms
for community detection available on the market.
Also, from a comparison among the various panels of
  Figure~\ref{fig:5}, it seems that 
the predictive power of the
exact recovery threshold deteriorates as the number of communities in
the stochastic model increases.

\section{Conclusions}
The analogy between the problems of identifying communities in networks
and communicating over a noisy channel is intrinsically present in all the 
methods for community detection based on statistical inference. 
In our contribution, we considered the analogy
explicitly as already done in Refs.~\cite{abbe2016exact, abbe2015community, 
abbe2017community}, and leveraged coding theory 
to achieve four main results. First, we built
a family of equivalent linear codes based on low-density parity-check (LDPC)
matrices, and show that they can be used to generate a class of 
LDPC community decoders. Second, we showed that the 
Shannon's noisy-channel coding
theorem sets a lower bound, named here as
decodability bound, on the maximal amount
of noise allowed for perfect community detection in 
the stochastic block model.
Third, we connected the first two results, 
showing that LDPC community decoders 
are potentially able to reach
the Shannon capacity of the stochastic block model.
Fourth, whereas the above results are valid for the simplest case 
of stochastic block models with two communities only, we also 
showed that the decodability bound can be easily extended to the case of
multiple communities providing a quantitatively accurate
prediction of the regimes of performance of the best
community detection algorithms available on 
the market ~\cite{lancichinetti2009community}.
This final result is certainly the most important from the
  practical point of view, as it seems to indicate that no much potential
  for improvement in performance is left.  
\change{We stress that this conclusion can, at the moment, 
be supported only by numerical evidence. 
This fact restricts ourselves to consider the conclusion valid
only for specific algorithms and specific settings of the stochastic block
model. We do not exclude that the best performing algorithms in the GN
and LFR benchmarks will fail in other models, 
as a recent theoretical study~\cite{peel2017ground}
demonstrated that a single algorithm outperforming all
other algorithms in every community detection 
task cannot exist. Further, as the mathematical proof of the Shannon's noisy-channel coding
theorem is valid only in the limit of infinitely-large systems, 
there is no mathematical guarantee that the decodability bound must
hold also for finite-size networks. Numerical evidence so far is
supportive, but, until a mathematical explanation is provided, 
there is always the chance to find an  algorithm able to beat the 
decodability bound.}

\acknowledgements{
The author acknowledges support from the National Science Foundation (Grant
CMMI-1552487), and from the US Army Research Office
(W911NF-16-1-0104).}



\appendix

\section{Spectral decoder}\label{app:1}

An algorithm that approximates a 
minimum distance decoder can be deployed as follows.
Our goal is to find the value of the information bits $\hat{\sigma}$
that lead to the minimum number of
violated parity-check equations [Eq.~(1) of the main text], given the 
word received. We can define a penalty
function as
\[
\begin{array}{ll}
D  = & \sum_{i>j}  A_{i,j}  (1 - \delta_{\sigma_i, \sigma_j}) + \sum_{i>j} (1 -
A_{i,j})  \delta_{\sigma_i, \sigma_j}   \\
& E - 2 \sum_{i>j}  A_{i,j}
\delta_{\sigma_i, \sigma_j}  + \sum_{i>j} \delta_{\sigma_i, \sigma_j}
\end{array}
 \;,
\]
where $\delta_{x,y} = 1$ if $x=y$, and $\delta_{x,y} = 0$ if $x \neq
y$, $A_{i,j}$ is the $i,j$ element of the adjacency matrix of the
received network, and $E$ is the total number of 
observed edges. 
Note that above sums are not performed in modulo $2$.
In the definition of the penalty function $D$, we
are not allowing for any correction on the parity bits received. Thus,
we can only act on the information bits.
The best estimates of the bits $\hat{\sigma}$
are such that $D$ is minimal. 
To approximate the solution, we can perform the 
transformation $\sigma_i = 0 \to \xi_i =1$,
and $\sigma_i = 1 \to \xi_i = -1$, so that $\delta_{\sigma_i, \sigma_j}
= (1 + \xi_i \xi_j)/2$, and
rewrite $D$ as
\[
\begin{array}{ll}
D  = & E -  \sum_{i>j}  A_{i,j} (1+\xi_i \xi_j)  + \frac{1}{2}\sum_{i>j}
(1+\xi_i \xi_j) \\
& \frac{N(N-1)}{4} + \frac{1}{2} \sum_{i>j} \xi_i \xi_j -
\sum_{i>j} A_{i,j} \xi_i \xi_j 
\end{array}
\; ,
\]
or in matrix-vector notation as
\begin{equation}
D = \frac{N(N-1)}{4} + \vec{\xi}^{\, T} (J/2 - \mathbb{I}/2  - A) \vec{\xi} \;, 
\label{eq:spectral}
\end{equation}
where $J$ is the all-one matrix, and $\mathbb{I}$ is the identity
matrix. The first term on the rhs of Eq.~(\ref{eq:spectral}) is
a constant dependent only on the size of the network $N$.
We need therefore to minimize only the 
rightmost term of the equation.
An approximate solution for the
configuration that minimizes $D$ can be found by
finding the largest eigenvector of the operator
$A - J/2 + \mathbb{I}/2$, and set $\hat{\sigma}_i =1$
if the corresponding component is smaller than zero, or $\hat{\sigma}_i
=0$, otherwise.

\section{Stochastic block model as a noisy channel}\label{app:llr}

In the stochastic block model, two nodes 
$i$ and $j$ belonging to the
same group, with corresponding parity bit $\theta_{i,j} = 0$, 
are connected with
probability $p_{in}$. The two nodes  $i$ and $j$  
belonging to different
groups , with corresponding
parity bit $\theta_{i,j} = 1$, are connected with
probability $p_{out}$. 
This means that parity bits
$\theta$ obey the rules of the
following binary asymmetric channel
\begin{equation}
\begin{array}{c|c|c}
\theta_{i,j} & A_{i,j} & P(A_{i,j}|\theta_{i,j})
\\
\hline
0 & 1 & p_{in}
\\
\hline
0 & 0 & 1 - p_{in}
\\
\hline
1 & 1 & p_{out}
\\
\hline
1 & 0 & 1 - p_{out}
\end{array} \; .
\label{eq:bac}
\end{equation}
We use the information about the noisy channel
to determine the best estimate of $\theta_{i,j}$ given the
observed value of $A_{i,j}$. We can write
\[
P(\theta_{i,j}|A_{i,j}) = \frac{P(A_{i,j} | \theta_{i,j})
  P(\theta_{i,j})} {P(A_{i,j})} \; .
\]
As we have no prior knowledge of the group assignments of the nodes,
and therefore about true values of the parity bit $\theta_{i,j}$, we can 
set $P(\theta_{i,j}) = 1/2$. Additionally, we can write
$P(A_{i,j} = 1) = (p_{in} + p_{out}) /2$ and $P(A_{i,j} = 0) = 1 -
(p_{in} + p_{out}) /2$. Using our knowledge of the noisy channel, we
can thus write
\[
P(\theta_{i,j} = 0|A_{i,j} = 1)  = \frac{p_{in}}{p_{in} + p_{out}}
\]
and
\[
P(\theta_{i,j} = 0 | A_{i,j} = 0 ) = \frac{1 - p_{in}}{2 - (p_{in} +
  p_{out})} \; .
\]
We can use those probabilities to determine the value
of the log-likelihood ratios (LLRs) 
\begin{widetext}
\begin{equation}
\ell_{i,j} = \log { \frac{ P(\theta_{i,j} = 0 | A_{i,j})  }  {
    P(\theta_{i,j} = 1 | A_{i,j})   }  } = \left\{
\begin{array}{ll}
\log{(p_{in})} - \log{(p_{out})} &  \textrm{ , if } A_{i,j} = 1
\\ 
\log{(1 - p_{in})} - \log{(1 - p_{out})} &  \textrm{ , if } A_{i,j} = 0
\end{array}
\right. \; .
\label{eq:llr_sbm}
\end{equation}
\end{widetext}

\section{Capacity of the channel}\label{app:capacity}

Messages are given by divisions of the network of $N$ nodes into
two blocks with size $n$ and $N-n$, respectively. Once $n$ is fixed, there will be
${N \choose n}$ equiprobable messages.
One of those messages
is encoded into a codeword $\vec{\theta}$ composed of
\[
\Theta_0 = {n \choose 2} + {N -n \choose 2} \; 
\]
parity bits equal to zero, and
\[
\Theta_1 = n (N-n) \; 
\]
parity bits equal to one. The 
length of the codeword is fixed and doesn't
depend on $n$
\[
L = \Theta_0 + \Theta_1 \; .
\]
Strictly speaking the codeword contains also information
bits. However, those bits are completely 
erased by the channel, thus for the computation of
the capacity, we can think that 
the codeword is composed of parity bits only.

In the stochastic block model, the relation between
transmitted and received parity bits is 
given by Eq.~(\ref{eq:bac}).  For simplicity, let us define
$\phi_{i,j} = (1 + A_{i,j}) \mod 2$.
As the probability that an individual received bit $\phi_{i,j}$ is
dependent only on the value of the 
bit transmitted $\theta_{i,j}$, we
can write
\[
P(\vec{\phi}|\vec{\theta}) = \prod_{(i,j)} P(\phi_{i,j}|\theta_{i,j}) \; .
\]

\begin{figure*}
\includegraphics[width = 0.6\textwidth]{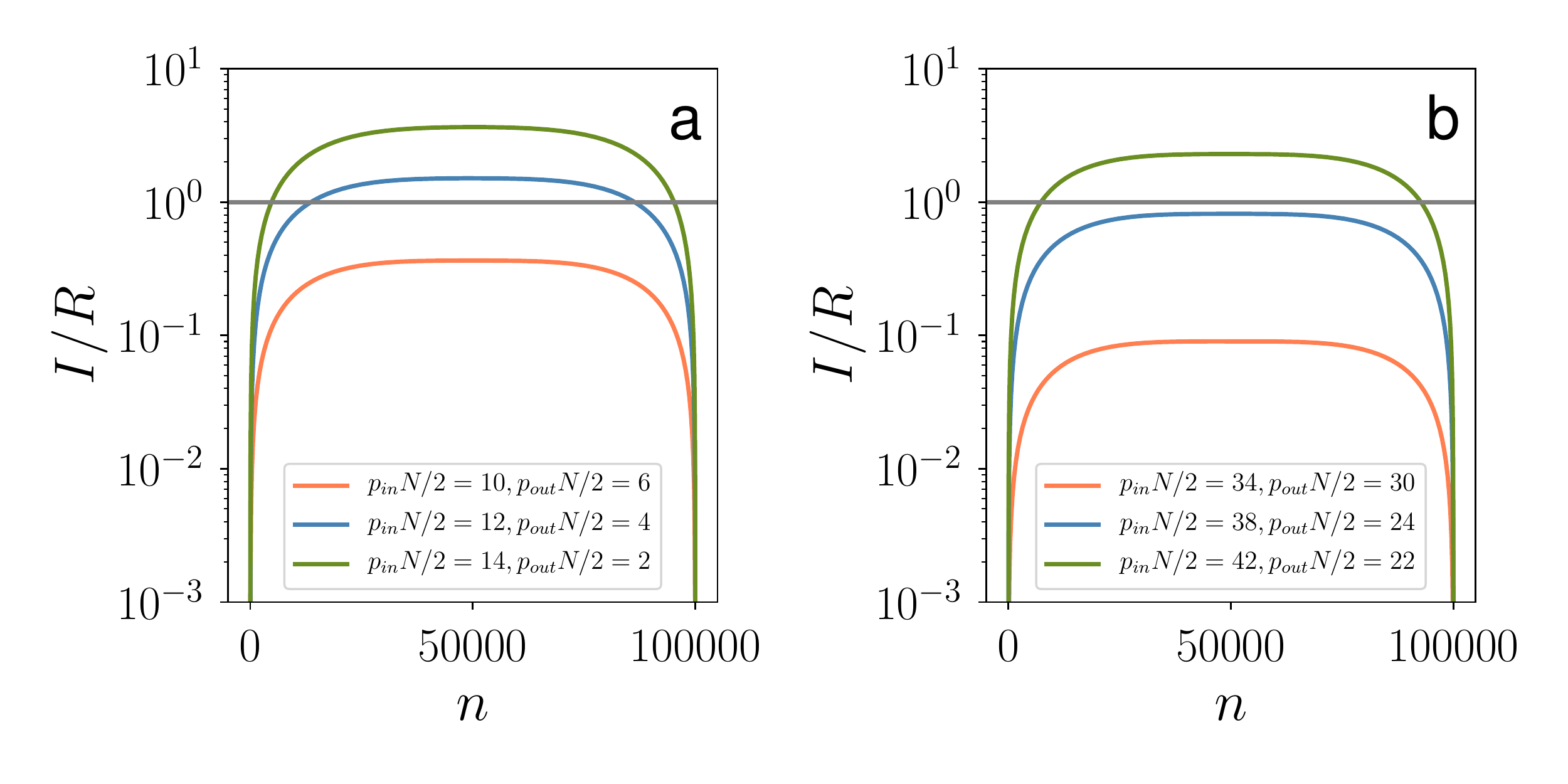}
\caption{(Color online) (a) Mutual information $I$ [Eq.~(\ref{eq:mi})] divided by the
  rate of the code $R = 2 / (N+2)$ as a function of the module size
  $n$  (the size of the other module is $N-n$). We consider fixed
  values of the probabilities $p_{in}$ and $p_{out}$.  Here, $N =
  10^5$. (b) Same as in panel a, but for different values of  $p_{in}$ and $p_{out}$.
}
\label{fig:capacity_sm2}
\end{figure*}

As the probability associated to the value of a generic received
parity 
bit $\phi$ is dependent only
on the value of the transmitted bit $\theta$, and for a fixed value of
$n$ all ${N \choose n}$ configurations are equiprobable, 
we can simply state that
\[
P(\theta =0|n) = \Theta_0 / (\Theta_0 + \Theta_1) = \alpha \;,
\]
and
\[
P(\theta =1|n) = \Theta_1 / (\Theta_0 + \Theta_1) = (1 - \alpha) \;.
\]
The parity bit $\phi$ is received as flipped with probability $p_{out}$ if
$\theta=1$, while it will stay equal to $\theta$ with probability
$p_{in}$ if $\theta=0$.
The conditional entropy is then given by
\[
H(\phi|\theta) = \alpha H_2(p_{in}) + (1-\alpha)
H_2(p_{out}) \; ,
\]
where
\[
H_2(f) = - f \log_2(f) - (1-f) \log_2(1-f)
\]
is the binary entropy function.
The probability that a received parity check bit $\phi$ is 
zero is
\[
P(\phi = 0 ) =  \alpha p_{in} + (1- \alpha) p_{out} \;.
\]
Thus, the entropy of the received bit $\phi$ is
\[
H(\phi) = H_2 [ \alpha p_{in} + (1- \alpha) p_{out} ] \;. 
\]
The expression for the mutual information reads
\begin{equation}
I(\theta;\phi) = H_2 [ \alpha p_{in} + (1- \alpha) p_{out} ] - \alpha H_2(p_{in}) - (1-\alpha)
H_2(p_{out}) \; .
\label{eq:mi}
\end{equation}
In Figure~\ref{fig:capacity_sm2}, we display the profile of the mutual information for specific settings of the stochastic block model.
To find the channel capacity, we need to maximize $I$ with
respect to $\alpha$. For simplicity, we will assume $\alpha$
continuous. The derivative of the binary entropy function is
\[
\frac{d}{dx} H_2(x) = \log_2 \left( x^{-1} - 1 \right) \;.
\]
The derivative of the mutual information is therefore
\[
\begin{array}{ll}
\frac{d}{d \alpha} I(\theta;\phi) = & ( p_{in} - p_{out}) \, \log_2 \left(
  \frac{1}{\alpha p_{in} + (1- \alpha) p_{out}}  - 1\right) 
\\ 
& - H_2(p_{in})
+  H_2(p_{out}) 
\end{array}
\; .
\]
Setting the previous expression equal to zero, we have
\[
( p_{in} - p_{out}) \, \log_2 \left(
  \frac{1}{\alpha^* p_{in} + (1- \alpha^*) p_{out}}  - 1\right) - H_2(p_{in})
+  H_2(p_{out})  = 0 
\]
\[
\log_2 \left(
  \frac{1}{\alpha^* p_{in} + (1- \alpha^*) p_{out}}  - 1\right) = \frac{H_2(p_{in})
-  H_2(p_{out}) } {p_{in} - p_{out}  }  
\]
\[
  \frac{1}{\alpha^* p_{in} + (1- \alpha^*) p_{out}}  - 1 = 2^{\frac{H_2(p_{in})
-  H_2(p_{out}) } {p_{in} - p_{out}  } } \;.
\]
For simplicity, let us define
\[
z = 2^{\frac{H_2(p_{in})
-  H_2(p_{out}) } {p_{in} - p_{out}  }}  \; \,
\]
thus 
\[
  \frac{1}{\alpha^* p_{in} + (1- \alpha^*) p_{out}}  = z + 1
\]
\[
\alpha^* p_{in} + (1- \alpha^*) p_{out} = \frac{1} {1 + z}
\]

\[
\alpha^*( p_{in} - p_{out}) = \frac{1 - p_{out} (1+ z)} {1 + z} 
\]
\[
\alpha^*= \frac{1 - p_{out} (1+ z)} {(1 + z) ( p_{in} - p_{out}) }  \;
,
\]
so that we have
\begin{equation}
\alpha^* = \frac{1 - p_{out} (1+ 2^{\frac{H_2(p_{in})
-  H_2(p_{out}) } {  p_{in} - p_{out}  }} )} {(1 + 2^{\frac{H_2(p_{in})
-  H_2(p_{out}) } {  p_{in} - p_{out}  }} ) (  p_{in} - p_{out}) } 
\label{eq:alpha}
\end{equation}
The capacity of the channel is 
\begin{equation}
C = H_2 [ \alpha^* p_{in} + (1- \alpha^*) p_{out} ] - \alpha^* H_2(p_{in}) - (1-\alpha^*)
H_2(p_{out}) \; .
\label{eq:capacity}
\end{equation}

\begin{figure*}
\includegraphics[width = 0.85\textwidth]{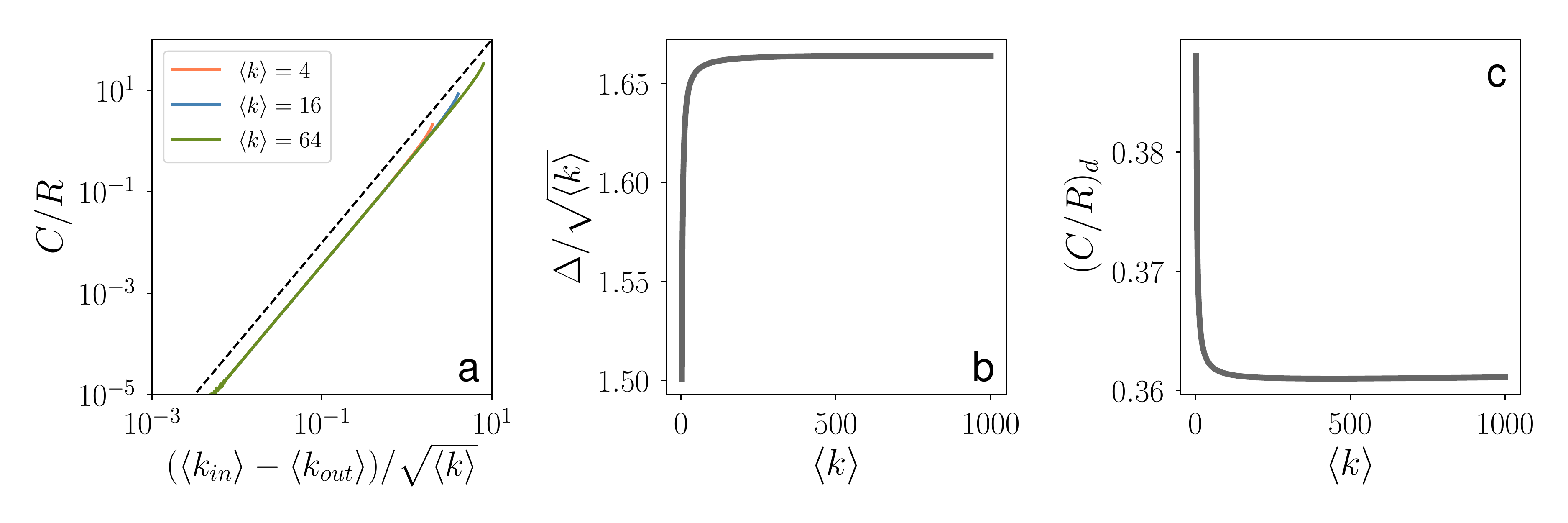}
\caption{(Color online)  (a) Relation between the ratio $C/R$, capacity divided by rate,
  and $(\langle k_{in} \rangle - \langle k_{out} \rangle) /
  \sqrt{\langle k \rangle
}$ for the stochastic block model with $N = 10^6$ and two groups
with identical size $n=N/2$. $\langle
k_{in} \rangle = p_{in} (n-1)$, $\langle
k_{out} \rangle = p_{out} n$, and $\langle k \rangle  = \langle k_{in}
\rangle + \langle k_{out} \rangle$. We consider different values of
$\langle k \rangle $.
The black dashed line identify is a power-law with exponent $2$.
(b) We compute $\Delta$ as the value of $\langle k_{in}
\rangle - \langle k_{out} \rangle$
for which $C/R =1$. As the plot shows, the value $\Delta /
\sqrt{\langle k \rangle}$
saturates quickly at $1.66$ as $\langle k \rangle$ grows.
(c) We determined the value of the ratio $C/R$ when $\langle k_{in}
\rangle - \langle k_{out}\rangle
= \sqrt{\langle k \rangle}$. We indicated this with $(C/R)_d$, and plotted the
quantity as a function of the average degree $\langle k \rangle$. We observe
a quick saturation $(C/R)_d \simeq 0.36$ as $\langle k \rangle$ grows.
}
\label{fig:capacity_sm1}
\end{figure*}

In Figure~\ref{fig:capacity_sm1}, we show that the ratio between channel capacity and code rate is a tight function of the parameters of the stochastic block model.

\section{Iterative decoding for the pair code}\label{app:pair}

We employ the Gallager algorithm to decode the received word.
The technique involves messages sent back and forth
between variable nodes to check nodes on the Tanner
graph constructed from the parity-check matrix $H$ of Eq.~(3) of the
main text.  We have a total of $N + N(N-1)/2$ variable nodes.
The first $N$ correspond to the actual nodes of the graph.
The other $N(N-1)/2$ variable nodes are instead given by all pairs
of nodes in the graph. Check nodes
amount to $N(N-1)/2$, each corresponding to a pair of nodes
in the graph. 

To describe variable nodes, we use the following
notation. The generic node $i$ has associated 
log-likelihood ratio (LLR)
\[
\ell_i = \log \, \frac{P(\sigma_i = 0 | s_i)}{P(\sigma_i = 1| s_i)} \;,
\]
where $s_i =0, 1$ is the received information bit, and $\sigma_i$ is the 
transmitted information bit. Please note that in our problem we
actually do not receive any information bit, as these are erased by
the channel.
Similarly, for a generic pair $(i,j)$ of nodes, we
define the LLR as
\[
\ell_{i,j} = \log \, \frac{P(\theta_{i,j}=0|A_{i,j})}{P(\theta_{i,j}=1|A_{i,j})} \;,
\]
where $\phi_{i,j} = 1 + A_{i,j}$ is the received parity bit.
Note that the former definition is perfectly symmetric
under the exchange of $i$ and $j$. It is clearly not defined, and
actually not used, for $i = j$. This fact is assumed below.

\

$\ell_i$ and $\ell_{i,j}$ are our best estimates of the value
of the variable nodes at stage $t=0$ of the algorithm.
At iteration $t$ of the algorithm, the variable node $i$ sends to the
check node $(i,j)$ the message 
\[
m^{(t)}_{i \to (i,j)} = \left\{
\begin{array}{ll}
\ell_i & \textrm{  , if } t=0
\\
\ell_i  + \sum_{k \neq j, k \neq i} n_{(i,k) \to i}^{(t-1)}  & \textrm{  , if } t
                                                    \geq 1
\end{array}
\right. \;.
\]
The message sent from the variable node $(i,j)$ to the check node
$(i,j)$ is instead equal to $\ell_{i,j}$ in all rounds of the algorithm.
In the above expression,  $n_{(i,j) \to i}$ is the message sent 
back from the check node  
$(i,j)$ to node $i$, and is defined as
\[
n^{(t)} _{(i,j) \to i} = \log \, \frac{1 + \, \tanh{(1/2 \, 
    m^{(t-1)} _{j \to (i,j)})}  \, \tanh{(1/2 \, 
    \ell_{i,j} )}   }  {1 - \, \tanh{(1/2 \, 
    m^{(t-1)} _{j \to (i,j)})}  \, \tanh{(1/2 \, 
    \ell_{i,j} )}   } \; .   
\]
The check node $(i,j)$ sends a message back also to the
variable node $(i,j)$ equal to
\[
n^{(t)} _{(i,j) \to (i,j)} = \log \, \frac{1 + \, \tanh{(1/2 \, 
    m^{(t-1)} _{i \to (i,j)})}  \, \tanh{(1/2 \, 
    m^{(t-1)} _{j \to (i,j)})}   }  {1 - \tanh{(1/2 \, 
    m^{(t-1)} _{i \to (i,j)})}  \, \tanh{(1/2 \, 
    m^{(t-1)} _{j \to (i,j)})}   } \; .   
\]
Please note that the latter message is not used in the iterative
algorithm.
It is however used to check the convergence of the algorithm as it follows.
At round $t>0$ of the algorithm, the estimated values of the LLRs are
\[
\begin{array}{ll}
\hat{\ell}^{(t)}_i = \ell_i  + \sum_{k \neq i} n_{(i,k) \to i}^{(t)} 
\\
\hat{\ell}^{(t)}_{i,j} = \ell_{i,j} + n^{(t)} _{(i,j) \to (i,j)}
\end{array} \; .
\]
The estimate of the bits associated with the variable
nodes is performed with a hard-decision choice, setting
$\hat{\sigma}_i = 0$ if $\ell^{(t)}_i<0$ and $\hat{\sigma}_i = 1$,
otherwise. Similarly for the best estimate of the pair variable 
$\hat{\theta}_{i,j} = 0$ if $\ell^{(t)}_{i,j}<0$ and
$\hat{\theta}_{i,j} = 1$, otherwise. Based on this choice, we can
establish if the bit string decoded at iteration $t$ is an actual
codeword, i.e., $(\hat{\sigma}_i + \hat{\sigma}_j +
\hat{\theta}_{i,j}) \mod 2 = 0$, for all $i \neq j$.
In such a case, we determine that the algorithm has converged.
Otherwise, we run the algorithm up to a desired maximal number of
iterations.

When applied to the stochastic block model, we can set $\ell_i = 0$
for all $i$, except for $\ell_{i^*} = \pm \infty$ for one node $i^*$.
The values of $\ell_{i,j}$ are instead provided in Eq.~(\ref{eq:llr_sbm}).

\subsection{Simplification of the decoding algorithm}

The structure of the equations above allows us to simplify the
decoding algorithm. As messages sent by pair-variables are 
unchanged, we can simply define
\[
\zeta^{(t)}_{i \to j} = m^{(t)}_{i \to (i,j)} \; ,
\]
to write $\zeta^{(t=0)}_{i \to j} = \ell_i$ and
\[
\zeta^{(t)}_{i \to j} = 
\ell_i  + \sum_{k \neq j}    \log \, \frac{1 + \, \tanh{(1/2 \, 
    \zeta^{(t-1)} _{k \to i}  )}  \, \tanh{(1/2 \, 
    \ell_{i,k}  )}   }  {1 - \, \tanh{(1/2 \, 
    \zeta^{(t-1)} _{k \to i}  )}  \, \tanh{(1/2 \, 
    \ell_{i,k}  )}  }   \;,
\]
for  $t \geq 1 $ .
The best estimates of the LLRs at stage $t \geq 1$ are
\[
\hat{\ell}^{(t)}_{i} = 
\ell_i  + \sum_{k}    \log \, \frac{1 + \, \tanh{(1/2 \, 
    \zeta^{(t-1)} _{k \to i}  )}  \, \tanh{(1/2 \, 
    \ell_{i,k}  )}   }  {1 - \, \tanh{(1/2 \, 
    \zeta^{(t-1)} _{k \to i}  )}  \, \tanh{(1/2 \, 
    \ell_{i,k}  )}  } 
\]
and
\[
\hat{\ell}_{i,j}^{(t)} = \ell_{i,j} +   \log \, \frac{1 + \, \tanh{(1/2 \, 
    \zeta^{(t-1)} _{j \to i}  )}  \, \tanh{(1/2 \, 
    \zeta^{(t-1)} _{i \to j}   )}   }  {1 - \, \tanh{(1/2 \, 
    \zeta^{(t-1)} _{j \to i}  )}  \, \tanh{(1/2 \, 
    \zeta^{(t-1)} _{i \to j}   )} } \; .
\]

\section{Iterative algorithm for the triplet code}\label{app:triplet}

If instead of the pair code, we consider the
triplet code

\[
\left( \theta_{i,j} + \theta_{i,k} + \theta_{j,k} \right) \mod 2 = 0
\; ,
\]

we can still use the Gallagher algorithm on the corresponding Tanner
graph. The Tanner graph contains $N(N-1)/2$ variable
nodes. Each of those variable nodes is connected to $N -2$ check
nodes. The total number of check nodes is $N(N-1)(N-2)/6$, each for
every triplet.

At iteration $t$ of the algorithm, the variable node $(i,j)$ sends to the
check node $(i,j,k)$ the message 
\[
m^{(t)}_{(i,j) \to (i,j,k)} = \left\{
\begin{array}{ll}
\ell_{i,j} & \textrm{  , if } t=0
\\
\ell_{i,j}  + \sum_{s \neq i, j, k} n_{(i,j,s) \to (i,j)}^{(t-1)}  & \textrm{  , if } t
                                                    \geq 1
\end{array}
\right. \;.
\]
The sum appearing above runs
over all triplets connected to $(i,j)$, excluding the triplet $(i,j,k)$.
In turn, check nodes reply to variable nodes with
\begin{widetext}
\[
n^{(t)} _{(i,j,k) \to (i,j)} = \log \, \frac{1 + \, \tanh{(1/2 \, 
    m^{(t-1)} _{(i,k) \to (i,j,k)})}  \, \tanh{(1/2 \, 
    m^{(t-1)} _{(j,k) \to (i,j,k)})}   }  {1 - \, \tanh{(1/2 \, 
    m^{(t-1)} _{(i,k) \to (i,j,k)})}  \, \tanh{(1/2 \, 
    m^{(t-1)} _{(j,k) \to (i,j,k)})}  } \; .   
\]
\end{widetext}
The reply depends only on the messages that the triplet $(i,j,k)$
received from the 
other two pairs attached to it, namely $(i,k)$ and $(j,k)$.
For $t \geq 1$, 
best estimates of the LLRs for variable nodes are:
\begin{widetext}
\[
\hat{\ell}^{(t)}_{i,j} = \ell_{i,j}  + \sum_{k} n_{(i,j,k) \to
  (i,j)}^{(t-1)}  = \ell_{i,j} + \sum_k  \log \, \frac{1 + \, \tanh{(1/2 \, 
    m^{(t-1)} _{(i,k) \to (i,j,k)})}  \, \tanh{(1/2 \, 
    m^{(t-1)} _{(j,k) \to (i,j,k)})}   }  {1 - \, \tanh{(1/2 \, 
    m^{(t-1)} _{(i,k) \to (i,j,k)})}  \, \tanh{(1/2 \, 
    m^{(t-1)} _{(j,k) \to (i,j,k)})}  }   \; .
\]
\end{widetext}

\section{Density evolution for the triplet code on the stochastic
  block model}\label{app:density}

For simplicity, we consider only the case of two equally sized
groups, so that $n = N/2$. 
Our plan is to monitor the evolution of the probability densities of the
log-likehood ratios (LLRs)
for internal and external pairs of nodes.
An internal pair of nodes consists in two nodes $i$ and $j$ within
the same group. We know that the true value of parity bit for such a
pair is $\theta_{i,j} = 0$. An external pair of nodes consists in two
nodes $i$ and $j$ 
belonging to different groups. The true value of the parity bit
associated
to this external pair is 
$\theta_{i,j} = 1$. 
At stage $t$ of the iterative algorithm, the LLR densities of external and internal pairs
are respectively indicated as $P_{in}^{(t)}(\hat{\ell})$ and
$P_{out}^{(t)}(\hat{\ell})$. These densities describe the behavior
of the LLRs over an infinite number of realizations of the stochastic
block model.
 To monitor the evolution of the LLR densities as functions 
of the iteration $t$ of the algorithm, we assume variables
to be independent. This assumption is correct up to $t=3$, as the girth
of the underlying Tanner graph is $6$. For a larger number of iterations,
variables in the true algorithm become dependent on each other, 
and they do not longer obey the 
distributions derived under the independence assumption.

\begin{figure}
\includegraphics[width = 0.45\textwidth]{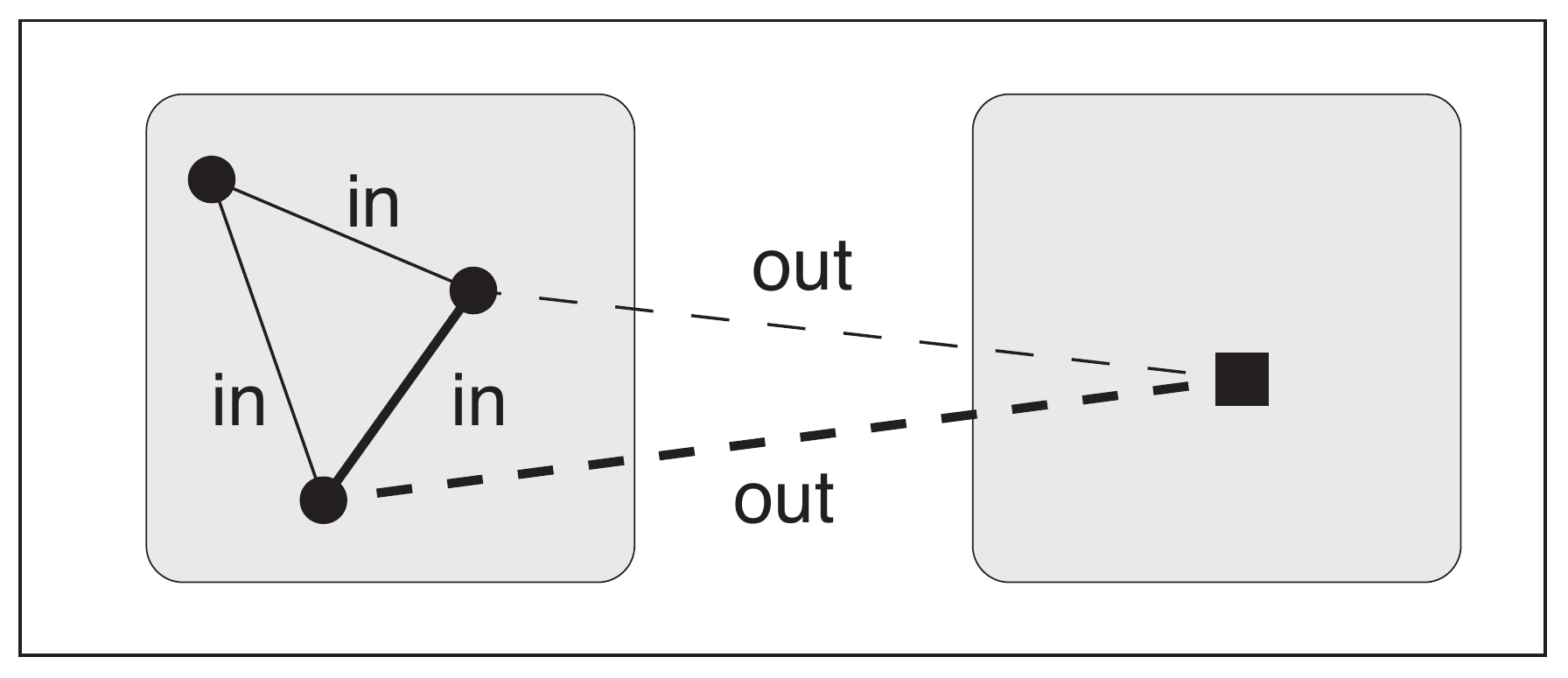}
\caption{(Color online)  Every internal pair of nodes is involved in two types of
  parity-check equations: (i) those formed with
other two internal pairs, and (ii) those formed with two external pairs. Every external
pairs of nodes is instead involved in parity-check equations with
one internal and one external pair.
}
\label{fig:illustration}
\end{figure}

For a generic internal pair, the initial value
of the LLR will be a random variable obeying the distribution
\[
P_{in}^{(t  = 0)}(\hat{\ell}) = \delta \left( \hat{\ell} - \log
  \frac{p_{in}}{p_{out}} \right) \,
p_{in} + \delta \left(\hat{\ell} - \log \frac{1 - p_{in}}{1- p_{out}} \right) (1 - p_{in}) \; ,
\]
where $\delta(x) = 1$ if $x =0$, and $\delta(x) =0$, otherwise.

For a generic external pair, the initial value
of the LLR is
a random variable obeying the distribution
\[
P_{out}^{(t  = 0)}(\hat{\ell}) = \delta \left( \hat{\ell} - \log
  \frac{p_{in}}{p_{out}} \right) \,
p_{out} + \delta \left(\hat{\ell} - \log \frac{1 - p_{in}}{1- p_{out}} \right) (1 - p_{out})\; .
\]

Every pair is connected to a total of $2n-2$ parity checks.
If the pair is internal, then the pair will be connected to $n-2$
parity checks that include other
two internal pairs, and $n$ parity checks that 
include two external pairs (see Fig.~\ref{fig:illustration}). For an external pair instead, all parity
checks necessarily include another external pair, and one
internal pair.

At iteration $t\geq 1$, the distribution of the LLR for a generic internal
pair is obtained as the sum of three independent contributions. 
The first term is a single random variable
extracted from $P_{in}^{(t = 0)}$, namely $z_{in}$.
The second term is given by the sum of $n-2$ random variables.
The value of each of these variables is obtained by first extracting two random
variables from $P_{in}^{(t  -1)}$, namely $x_{in}^{(g)}$ and
$y_{in}^{(g)}$, and
then computing the quantity 
\[
q^{(g)}_{in, in} = \log \frac{ 1 + \tanh (x_{in}^{(g)}/2)
  \tanh (y_{in}^{(g)}/2)   }{ 1 - \tanh (x_{in}^{(g)}/2)
  \tanh (y_{in}^{(g)}/2)   } \; .
\]
The value of the second term is given by
\[
q_{in, in} = \sum_{g=1}^{n-2} q^{(g)}_{in, in} \;.
\]
The third term is given by the sum of $n$ random variables, generated
from the sum of two random variables $x_{out}^{(g)}$ and
$y_{out}^{(g)}$,
extracted at random from the distribution $P_{out}^{(t  -1)}$.
For a given pair of random variables, we compute
\[
q^{(g)}_{out, out} = \log \frac{ 1 + \tanh (x_{out}^{(g)}/2) \tanh (y_{out}^{(g)}/2)    }{ 1 - \tanh (x_{out}^{(g)}/2) \tanh (y_{out}^{(g)}/2)   } \; .
\]
The value of the third term is finally given by
\[
q_{out, out} = \sum_{g=1}^{n} q^{(g)}_{out, out} \;.
\]

As the various quantities are determined independently, 
the distribution of the LLR for internal pairs after the $t$-th
iteration is
\begin{equation}
\begin{array}{ll}
P_{in}^{(t )}(\hat{\ell}) = & P^{(t=0)}(z_{in}) P^{(t-1)}(q_{in, in}) 
\\
& \times \, P^{(t-1)}(q_{out, out}) \; \delta( \hat{\ell} -
z_{in} - q_{in, in}- q_{out, out}   ) 
\end{array}
\; , 
\label{eq:llr_in}
\end{equation}

where $P^{(t=0)}(z_{in})$,  $P^{(t-1)}(q_{in, in})$,  and $P^{(t-1)}(q_{out, out})$ are
respectively the
probability distributions of the variables $z_{in}$,  $q_{in, in}$,
and  $q_{out, out}$ as defined above.

For external pairs, the computation of the distribution of the LLRs is
very similar. There will be two contributions. The first is just a
random variable extracted from $P_{out}^{(t  = 0)}$, namely $z_{out}$.
The second is computed by extracting two random numbers $x_{in}^{(g)}$
and  $y_{out}^{(g)}$, respectively from the distributions $P_{in}^{(t  -1)}$
and $P_{out}^{(t  -1
  )}$. One then evaluates the quantity
\[
q^{(g)}_{in, out} = \log \frac{ 1 + \tanh (x_{in}^{(g)}/2) \tanh (y_{out}^{(g)}/2)   }{ 1 - \tanh (x_{in}^{(g)}/2) \tanh (y_{out}^{(g)}/2)   } \; .
\]
The value of the second term is finally given by
\[
q_{in, out} = \sum_{g=1}^{2n-2} q^{(g)}_{in, out} \; ,
\]
and the distribution of the LLR for external pairs is given by

\begin{equation}
P_{out}^{(t )}(\hat{\ell}) = P^{(t = 0)}(z_{out}) P^{(t-1)}(q_{in, out})  \;  \delta( \hat{\ell} -
z_{out} -  q_{in, out}   ) \;.
\label{eq:llr_out}
\end{equation}

\subsection{Approximation for large networks}

For $N \gg 1$, we expect that
the distributions $P^{(t)}(q_{in, in})$,  $P^{(t)}(q_{out, out})$, and $P^{(t)}(q_{in, out})$
appearing at iterations $t \geq 1$
are well described by Normal distributions, so that
\[
P^{(t)}(q_{in, in}) \simeq \mathcal{N} (q_{in, in};  (n-2) \mu^{(t)}_{in, in},  \sqrt{n-2}
\sigma^{(t)}_{in, in} ) \; ,
\]
\[
P^{(t)}(q_{out, out}) \simeq \mathcal{N} ( q_{out, out} ; n \mu^{(t)}_{out, out},  \sqrt{n}
\sigma^{(t)}_{out, out} )
\]
and
\[
P^{(t)}(q_{in, out}) \simeq \mathcal{N} (q_{in, out};  (2n-2) \mu^{(t)}_{in, out},  \sqrt{2n-2}
\sigma^{(t)}_{in, out} ) \;. 
\]
We used here $\mathcal{N} (x; \mu, \sigma )$ to indicate that the
variable $x$ is distributed according to a normal distribution withe
average $\mu$ and standard deviation $\sigma$.

If we define
\[
g(x,y) = \log \frac{ 1 + \tanh (x/2) \tanh (y/2)   }{ 1 - \tanh
  (x/2) \tanh (y/2)   }
\]

We have
\[
\mu^{(t)}_{in, in} = \int dx \int dy  \; P^{(t)}_{in}(x) \, P^{(t)}_{in}(y)  \,
 \, g(x, y) 
\, 
\]
and
\[
(\sigma^{(t)}_{in, in})^2 + (\mu^{(t)}_{in, in})^2 = \int dx \int dy  \; P^{(t)}_{in}(x) \, P^{(t)}_{in}(y) \,
 [ g(x, y) ]^2 
\, .
\]

Similar expressions can be written for $\mu^{(t)}_{out, out}$, $\sigma^{(t)}_{out,
  out}$, $\mu^{(t)}_{in, out}$, and $\sigma^{(t)}_{in, out}$.
The updated values of the distributions $
P^{(t+1)}_{in}(\hat{\ell})$ and $P^{(t+1)}_{out}(\hat{\ell})$ for stage $t+1$ of
the algorithm are given by 
\begin{widetext}
\begin{equation}
P^{(t +1)}_{in}(\hat{\ell})  = p_{in} \,   \mathcal{N} \{ \hat{\ell};
  \mu^{(t)}_{in} + \log{(p_{in}/p_{out})}, \sigma^{(t)}_{in}  \} +  (1-p_{in}) \,
  \mathcal{N} \{ \hat{\ell};
  \mu^{(t)}_{in} + \log{[(1-p_{in})/(1-p_{out})]}, \sigma^{(t)}_{in}
  \}
\label{eq:llr_in1}
\end{equation}
\end{widetext}
and
\begin{widetext}
\begin{equation}
P^{(t +1)}_{out}(\hat{\ell})  = p_{out} \,   \mathcal{N} \{ \hat{\ell};
  \mu^{(t)}_{out} + \log{(p_{in}/p_{out})}, \sigma^{(t)}_{out}  \} +  (1-p_{out}) \,
  \mathcal{N} \{ \hat{\ell};
  \mu^{(t)}_{out} + \log{[(1-p_{in})/(1-p_{out})]}, \sigma^{(t)}_{out}  \} \; ,
\label{eq:llr_out1}
\end{equation}
\end{widetext}
where 
\[\mu^{(t)}_{in} = (n-2) \mu^{(t)}_{in, in} + n \mu^{(t)}_{out, out}
  \; ,
\] 
\[
(\sigma^{(t)}_{in})^2 = (n-2) (\sigma^{(t)}_{in, in})^2 + n (\sigma^{(t)}_{out, out})^2
\; , 
\] 
\[
\mu^{(t)}_{out} = (2n-2) \mu^{(t)}_{in, out} \; ,
\] 
and
\[
(\sigma^{(t)}_{out})^2 = (2n-2) (\sigma^{(t)}_{in,
  out})^2 \; .
\]
Eqs.~(\ref{eq:llr_in1}) and Eqs.~(\ref{eq:llr_out1}) follow
directly
from Eqs.~(\ref{eq:llr_in}) and Eqs.~(\ref{eq:llr_out}), as the
distribution involved in the convolution are only normal and delta distributions.
Eqs. ~(\ref{eq:llr_in1}) and Eqs.~(\ref{eq:llr_out1}) allow us to
compute the probability of bit error as

\begin{equation}
p^{(t)}_e = \frac{1}{2} \, (\epsilon_{in} + \epsilon_{out} ) \; ,
\label{eq:err}
\end{equation}
where we used the approximation $\alpha \simeq 1/2$ for sufficiently
large values of $N$, and
\[
\epsilon_{in} = \int_{- \infty}^0 \, d\ell \; 
P^{(t)}_{in}(\ell)  
\]
and
\[
\epsilon_{out} = \int_{0}^{+ \infty} \, d\ell \; 
P^{(t)}_{out}(\ell) \;.
\]
We can further estimate the probability that one
randomly chosen parity-check equation is violated as
\begin{widetext}
\begin{equation}
p^{(t)}_s = 1 - \frac{ w_{in}   [  (1 - \epsilon_{in})^3 + 3 \epsilon_{in}^2 (1
  - \epsilon_{in} ) ]  + w_{out}   [  (1 - \epsilon_{in}) 
  \epsilon_{out}^2  +  2 \epsilon_{in} \epsilon_{out} (1
  - \epsilon_{out} )  + (1 - \epsilon_{in}) \epsilon_{out}^2 ] } {w_{in} + w_{out} } \; ,
\label{eq:err_eq}
\end{equation}
\end{widetext}
where
\[
w_{in} = \frac{n (n-1)(n-2)}{3} \; ,
\]
and
\[
w_{in} = n^2 (n-1) \; .
\]

\begin{figure*}
\includegraphics[width = 0.9\textwidth]{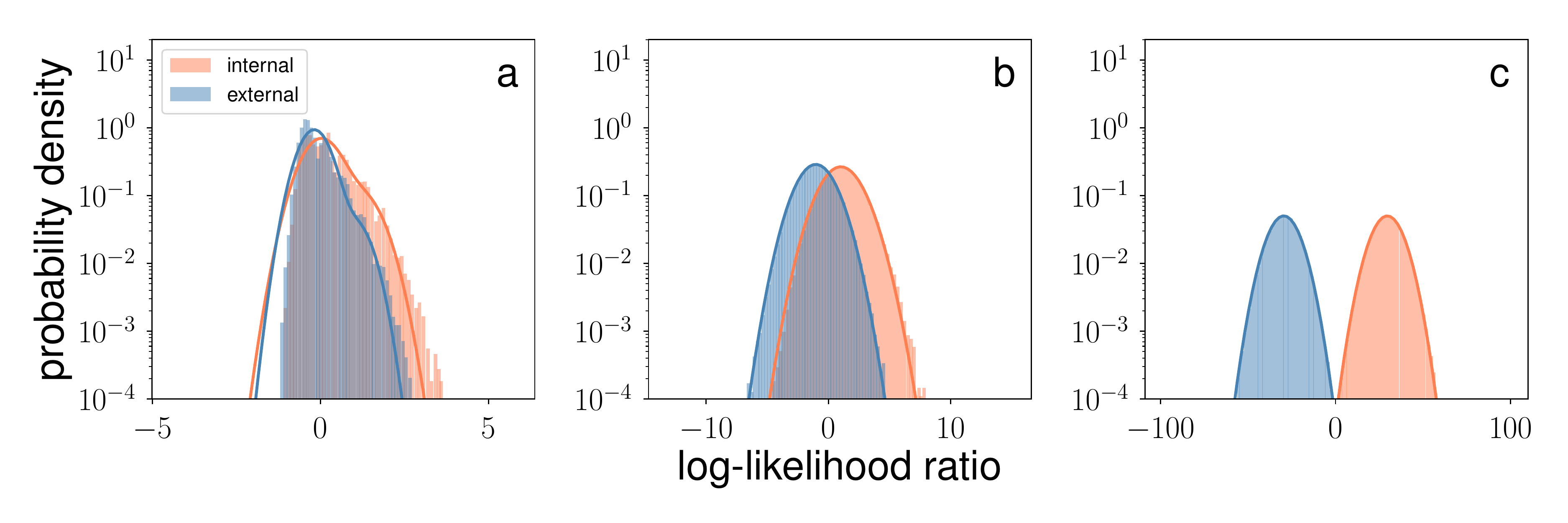}
\caption{(Color online)  Density distribution $P_{in}^{(t)}(\hat{\ell})$ of the log-likelihood ratio for internal
pairs (orange), and  density distribution $P_{out}^{(t)}(\hat{\ell})$ of the log-likelihood ratio for external
pairs (blue) for a network with $N = 100$ nodes, and equally sized
groups with
$n=50$ nodes. We consider $(n-1) p_{in} = \langle k_{in} \rangle = 6$, and
$n p_{out} = \langle k_{out} \rangle = 2$. The plot shows how the
distributions change as functions of the iteration $t$ of the
algorithm: (a) $t=1$, (b) $t=2$, and (c) $t=3$. Results of numerical
simulations (bars) are compared with the normal approximations (lines)
of Eqs.~(\ref{eq:llr_in1}) and ~(\ref{eq:llr_out1}).}
\label{fig:dens_sm1}
\end{figure*}

\begin{figure*}
\includegraphics[width = 0.9\textwidth]{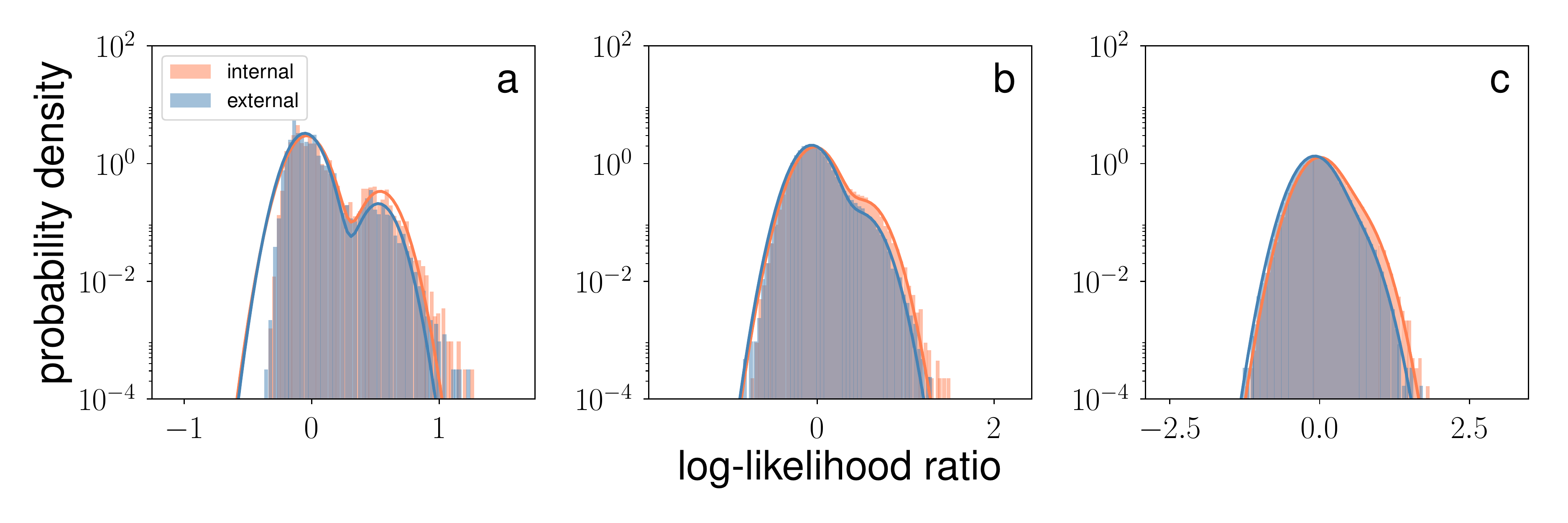}
\caption{(Color online)  Same as in Fig.~\ref{fig:dens_sm1} but for $(n-1) p_{in} = \langle k_{in} \rangle = 5$, and
$n p_{out} = \langle k_{out} \rangle = 3$.}
\label{fig:dens_sm2}
\end{figure*}


\section{Capacity of stochastic block models with more than two
  groups}
So far, we considered the simplest scenario of stochastic block models
composed of only two groups. Under this assumption, the problem 
of identifying the memberships of the groups can be
mapped into a decoding task of a linear code. Writing linear codes
that apply to scenarios where the presence of multiple groups is
allowed 
seems challenging. However, we can 
still provide insights to the problem by simply studying the channel
characteristics, and relying on the Shannon theorem to provide
indications
about the regime of exact recovery. The non-constructive nature
of the Shannon theorem itself allows us to draw conclusions
directly from the rate of the code and the channel features
without the need of necessarily specifying a encoding/decoding protocol.
From the graphical point of view,
the situation of multiple groups is identical to the one of two
groups. Still, we can immagine that the encoder generates a network of
disconnected cliques depending on the memberships of the nodes, and
that the channel flips the values of those  bits according to
some stochastic rule. 

Indicate with $Q$ the total number of groups in the networks, and with
$n_i$ the number of nodes in group $i$. We clearly have that
\[
\sum_{i=1}^K n_i = N \; .
\]
The total number of parity bits equal to zero in the transmitted 
codeword is
\[
\Theta_0 = \sum_{i=1}^{Q}  \frac{n_i(n_i -1)}{2} = \frac{1}{2} 
\sum_{i=1}^Q n_i^2 - \frac{N}{2}  \; .
\]
The total number of parity bits equal to one is instead
\[
\Theta_1 = \frac{N(N-1)}{2} - \Theta_0 \; .
\] 
Exactly as in the case of two groups, it is convenient to define
\[
\alpha = \frac{\Theta_0}{\Theta_0 + \Theta_1}
\]
as the relative amount of parity bits equal to zero.
The rate of the code is given in Eq.~(\ref{eq:rate_multi}).

For simplicity, let us focus on the case in
which the channel acts on the codeword using 
the rules of a binary asymmetric channel [Eq.~(\ref{eq:bac})].
Under these conditions, everything we wrote for the case of two groups
up to Eq.~(\ref{eq:mi}) still holds.
The fundamental difference here is that $\alpha$ 
depends on the size of the $Q$ groups. In principle, the maximization
of the mutual information requires to take derivatives with respect
to $Q-1$ variables. The profile of 
mutual information in this hyper-dimensional space is similar
to the one appearing in Fig.~\ref{fig:capacity_sm2}, looking flat over
a big number of different configurations. This fact allows us to
assume that the maximum of the mutual information is also reached for
equally sized groups, $n_i = N/Q$ for all $i$, so that we can write
\[
\Theta_0^* = \frac{N (N/Q -1) }{2} \; ,
\]
and
\begin{equation}
\alpha^* = \frac{N/Q-1}{N-1} \; .
\label{eq:alpha_multi_app}
\end{equation}
Finally, we can obtain the channel capacity as
\begin{equation}
C = H_2 [ \alpha^* p_{in} + (1- \alpha^*) p_{out} ] - \alpha^* H_2(p_{in}) - (1-\alpha^*)
H_2(p_{out}) 
\; .
\label{eq:cap_multi}
\end{equation}

\section{Comparison with the exact recovery threshold}\label{app:abbe}

In Refs.~\cite{abbe2016exact, abbe2017community}, the authors provided 
an exact estimate
of the threshold that the parameters of the stochastic block model
must satisfy to have necessary and sufficient conditions for exact
recovery of the modules.
We report here only the results for 
the symmetric case, where a network with $N$ nodes is
divided into $Q$ groups of size $N/Q$.
Pairs of nodes in the same group are connected with probability
$p_{in}$, whereas pairs of nodes in different groups are connected
with
probability $p_{out}$.
The condition reads
\begin{equation}
(\sqrt{a} - \sqrt{b} )^2 = Q\; ,
\label{eq:abbe1}
\end{equation}
where 
\[
a = \frac{p_{in} N}{\log{N}} \qquad \textrm{ and }
\qquad b =
\frac{p_{out} N}{\log{N}} \; .
\]
As in our analysis we often consider the decodability of the
model as a function of the difference between
average internal and external degree and fixed average total degree,
in the following we rewrite Eq.~(\ref{eq:abbe1}) in these terms.
We note that
\[
a = \frac{Q \, \langle k_{in} \rangle } {\log N} \qquad \textrm{ and }
\qquad b = \frac{Q \, \langle k_{out}  \rangle } {(Q-1) \, \log N} \;.
\]
The solution of Eq.~(\ref{eq:abbe1}) is
\[
b^* = \frac{c \pm \sqrt{2cQ - Q^2}}{2} \;,
\]
where 
\[
c = \frac{Q \, \langle k_{in} \rangle + Q/(Q-1) \, \langle k_{out} \rangle}{\log N}\; .
\]
The value of $a$ at threshold is
\[
a^* = c - b^* \; .
\]
For the case $Q=2$, we can write the threshold as
\begin{equation}
|\langle k_{in} \rangle   - \langle k_{out} \rangle|   = \log N \, \sqrt{ \frac{2 \langle
    k \rangle}{\log N} - 1  }  \; .
\label{eq:abbe2a}
\end{equation}
For $Q>2$, the value of the mixing 
parameter~\cite{lancichinetti2009community}
for which perfect recovery is allowed is given by
\begin{equation}
\mu^* = 
\frac{(Q-1) \, \log N \,  b^* } {\langle k \rangle \, Q }\; .
\label{eq:abbe2b}
\end{equation}
The condition for exact recovery is found imposing $\mu^* = \mu$.


\begin{thebibliography}{37}
\expandafter\ifx\csname natexlab\endcsname\relax\def\natexlab#1{#1}\fi
\expandafter\ifx\csname bibnamefont\endcsname\relax
  \def\bibnamefont#1{#1}\fi
\expandafter\ifx\csname bibfnamefont\endcsname\relax
  \def\bibfnamefont#1{#1}\fi
\expandafter\ifx\csname citenamefont\endcsname\relax
  \def\citenamefont#1{#1}\fi
\expandafter\ifx\csname url\endcsname\relax
  \def\url#1{\texttt{#1}}\fi
\expandafter\ifx\csname urlprefix\endcsname\relax\def\urlprefix{URL }\fi
\providecommand{\bibinfo}[2]{#2}
\providecommand{\eprint}[2][]{\url{#2}}

\bibitem[{\citenamefont{Fortunato}(2010)}]{fortunato2010community}
\bibinfo{author}{\bibfnamefont{S.}~\bibnamefont{Fortunato}},
  \bibinfo{journal}{Phys. Rep.} \textbf{\bibinfo{volume}{486}},
  \bibinfo{pages}{75} (\bibinfo{year}{2010}).

\bibitem[{\citenamefont{Girvan and Newman}(2002)}]{girvan2002community}
\bibinfo{author}{\bibfnamefont{M.}~\bibnamefont{Girvan}} \bibnamefont{and}
  \bibinfo{author}{\bibfnamefont{M.~E.} \bibnamefont{Newman}},
  \bibinfo{journal}{Proc. Natl. Acad. Sci. U.S.A.}
  \textbf{\bibinfo{volume}{99}}, \bibinfo{pages}{7821} (\bibinfo{year}{2002}).

\bibitem[{\citenamefont{Radicchi et~al.}(2004)\citenamefont{Radicchi,
  Castellano, Cecconi, Loreto, and Parisi}}]{radicchi2004defining}
\bibinfo{author}{\bibfnamefont{F.}~\bibnamefont{Radicchi}},
  \bibinfo{author}{\bibfnamefont{C.}~\bibnamefont{Castellano}},
  \bibinfo{author}{\bibfnamefont{F.}~\bibnamefont{Cecconi}},
  \bibinfo{author}{\bibfnamefont{V.}~\bibnamefont{Loreto}}, \bibnamefont{and}
  \bibinfo{author}{\bibfnamefont{D.}~\bibnamefont{Parisi}},
  \bibinfo{journal}{Proc. Natl. Acad. Sci. U.S.A.}
  \textbf{\bibinfo{volume}{101}}, \bibinfo{pages}{2658} (\bibinfo{year}{2004}).

\bibitem[{\citenamefont{Newman}(2013)}]{newman2013spectral}
\bibinfo{author}{\bibfnamefont{M.~E.} \bibnamefont{Newman}},
  \bibinfo{journal}{Phys. Rev. E} \textbf{\bibinfo{volume}{88}},
  \bibinfo{pages}{042822} (\bibinfo{year}{2013}).

\bibitem[{\citenamefont{Newman and Girvan}(2004)}]{newman2004finding}
\bibinfo{author}{\bibfnamefont{M.~E.} \bibnamefont{Newman}} \bibnamefont{and}
  \bibinfo{author}{\bibfnamefont{M.}~\bibnamefont{Girvan}},
  \bibinfo{journal}{Phys. Rev. E} \textbf{\bibinfo{volume}{69}},
  \bibinfo{pages}{026113} (\bibinfo{year}{2004}).

\bibitem[{\citenamefont{Reichardt and
  Bornholdt}(2004)}]{reichardt2004detecting}
\bibinfo{author}{\bibfnamefont{J.}~\bibnamefont{Reichardt}} \bibnamefont{and}
  \bibinfo{author}{\bibfnamefont{S.}~\bibnamefont{Bornholdt}},
  \bibinfo{journal}{Phys. Rev. Lett.} \textbf{\bibinfo{volume}{93}},
  \bibinfo{pages}{218701} (\bibinfo{year}{2004}).

\bibitem[{\citenamefont{Rosvall and Bergstrom}(2007)}]{rosvall2007information}
\bibinfo{author}{\bibfnamefont{M.}~\bibnamefont{Rosvall}} \bibnamefont{and}
  \bibinfo{author}{\bibfnamefont{C.~T.} \bibnamefont{Bergstrom}},
  \bibinfo{journal}{Proc. Natl. Acad. Sci. U.S.A.}
  \textbf{\bibinfo{volume}{104}}, \bibinfo{pages}{7327} (\bibinfo{year}{2007}).

\bibitem[{\citenamefont{Rosvall and Bergstrom}(2008)}]{rosvall2008maps}
\bibinfo{author}{\bibfnamefont{M.}~\bibnamefont{Rosvall}} \bibnamefont{and}
  \bibinfo{author}{\bibfnamefont{C.~T.} \bibnamefont{Bergstrom}},
  \bibinfo{journal}{Proc. Natl. Acad. Sci. U.S.A.}
  \textbf{\bibinfo{volume}{105}}, \bibinfo{pages}{1118} (\bibinfo{year}{2008}).

\bibitem[{\citenamefont{Decelle
  et~al.}(2011{\natexlab{a}})\citenamefont{Decelle, Krzakala, Moore, and
  Zdeborov{\'a}}}]{decelle2011inference}
\bibinfo{author}{\bibfnamefont{A.}~\bibnamefont{Decelle}},
  \bibinfo{author}{\bibfnamefont{F.}~\bibnamefont{Krzakala}},
  \bibinfo{author}{\bibfnamefont{C.}~\bibnamefont{Moore}}, \bibnamefont{and}
  \bibinfo{author}{\bibfnamefont{L.}~\bibnamefont{Zdeborov{\'a}}},
  \bibinfo{journal}{Phys. Rev. Lett.} \textbf{\bibinfo{volume}{107}},
  \bibinfo{pages}{065701} (\bibinfo{year}{2011}{\natexlab{a}}).

\bibitem[{\citenamefont{Karrer and Newman}(2011)}]{karrer2011stochastic}
\bibinfo{author}{\bibfnamefont{B.}~\bibnamefont{Karrer}} \bibnamefont{and}
  \bibinfo{author}{\bibfnamefont{M.~E.} \bibnamefont{Newman}},
  \bibinfo{journal}{Phys. Rev. E} \textbf{\bibinfo{volume}{83}},
  \bibinfo{pages}{016107} (\bibinfo{year}{2011}).

\bibitem[{\citenamefont{Peixoto}(2014)}]{peixoto2014hierarchical}
\bibinfo{author}{\bibfnamefont{T.~P.} \bibnamefont{Peixoto}},
  \bibinfo{journal}{Phys. Rev. X} \textbf{\bibinfo{volume}{4}},
  \bibinfo{pages}{011047} (\bibinfo{year}{2014}).

\bibitem[{\citenamefont{Peixoto}(2013)}]{peixoto2013parsimonious}
\bibinfo{author}{\bibfnamefont{T.~P.} \bibnamefont{Peixoto}},
  \bibinfo{journal}{Phys. Rev. Lett.} \textbf{\bibinfo{volume}{110}},
  \bibinfo{pages}{148701} (\bibinfo{year}{2013}).

\bibitem[{\citenamefont{Peixoto}(2017)}]{peixoto2017bayesian}
\bibinfo{author}{\bibfnamefont{T.~P.} \bibnamefont{Peixoto}},
  \bibinfo{journal}{arXiv preprint arXiv:1705.10225}  (\bibinfo{year}{2017}).

\bibitem[{\citenamefont{Moore}(2017)}]{moore2017computer}
\bibinfo{author}{\bibfnamefont{C.}~\bibnamefont{Moore}},
  \bibinfo{journal}{arXiv preprint arXiv:1702.00467}  (\bibinfo{year}{2017}).

\bibitem[{\citenamefont{Nadakuditi and Newman}(2012)}]{nadakuditi2012graph}
\bibinfo{author}{\bibfnamefont{R.~R.} \bibnamefont{Nadakuditi}}
  \bibnamefont{and} \bibinfo{author}{\bibfnamefont{M.~E.}
  \bibnamefont{Newman}}, \bibinfo{journal}{Phys. Rev. Lett.}
  \textbf{\bibinfo{volume}{108}}, \bibinfo{pages}{188701}
  (\bibinfo{year}{2012}).

\bibitem[{\citenamefont{Krzakala et~al.}(2013)\citenamefont{Krzakala, Moore,
  Mossel, Neeman, Sly, Zdeborov{\'a}, and Zhang}}]{krzakala2013spectral}
\bibinfo{author}{\bibfnamefont{F.}~\bibnamefont{Krzakala}},
  \bibinfo{author}{\bibfnamefont{C.}~\bibnamefont{Moore}},
  \bibinfo{author}{\bibfnamefont{E.}~\bibnamefont{Mossel}},
  \bibinfo{author}{\bibfnamefont{J.}~\bibnamefont{Neeman}},
  \bibinfo{author}{\bibfnamefont{A.}~\bibnamefont{Sly}},
  \bibinfo{author}{\bibfnamefont{L.}~\bibnamefont{Zdeborov{\'a}}},
  \bibnamefont{and} \bibinfo{author}{\bibfnamefont{P.}~\bibnamefont{Zhang}},
  \bibinfo{journal}{Proc. Natl. Acad. Sci. U.S.A.}
  \textbf{\bibinfo{volume}{110}}, \bibinfo{pages}{20935}
  (\bibinfo{year}{2013}).

\bibitem[{\citenamefont{Radicchi}(2013)}]{radicchi2013detectability}
\bibinfo{author}{\bibfnamefont{F.}~\bibnamefont{Radicchi}},
  \bibinfo{journal}{Phys. Rev. E} \textbf{\bibinfo{volume}{88}},
  \bibinfo{pages}{010801} (\bibinfo{year}{2013}).

\bibitem[{\citenamefont{Radicchi}(2014)}]{radicchi2014paradox}
\bibinfo{author}{\bibfnamefont{F.}~\bibnamefont{Radicchi}},
  \bibinfo{journal}{EPL} \textbf{\bibinfo{volume}{106}}, \bibinfo{pages}{38001}
  (\bibinfo{year}{2014}).

\bibitem[{\citenamefont{Young et~al.}(2017)\citenamefont{Young, Desrosiers,
  H\'ebert-Dufresne, Laurence, and Dub\'e}}]{PhysRevE.95.062304}
\bibinfo{author}{\bibfnamefont{J.-G.} \bibnamefont{Young}},
  \bibinfo{author}{\bibfnamefont{P.}~\bibnamefont{Desrosiers}},
  \bibinfo{author}{\bibfnamefont{L.}~\bibnamefont{H\'ebert-Dufresne}},
  \bibinfo{author}{\bibfnamefont{E.}~\bibnamefont{Laurence}}, \bibnamefont{and}
  \bibinfo{author}{\bibfnamefont{L.~J.} \bibnamefont{Dub\'e}},
  \bibinfo{journal}{Phys. Rev. E} \textbf{\bibinfo{volume}{95}},
  \bibinfo{pages}{062304} (\bibinfo{year}{2017}).

\bibitem[{\citenamefont{Danon et~al.}(2005)\citenamefont{Danon, Diaz-Guilera,
  Duch, and Arenas}}]{danon2005comparing}
\bibinfo{author}{\bibfnamefont{L.}~\bibnamefont{Danon}},
  \bibinfo{author}{\bibfnamefont{A.}~\bibnamefont{Diaz-Guilera}},
  \bibinfo{author}{\bibfnamefont{J.}~\bibnamefont{Duch}}, \bibnamefont{and}
  \bibinfo{author}{\bibfnamefont{A.}~\bibnamefont{Arenas}},
  \bibinfo{journal}{J. Stat. Mech. Theory Exp.}
  \textbf{\bibinfo{volume}{2005}}, \bibinfo{pages}{P09008}
  (\bibinfo{year}{2005}).

\bibitem[{\citenamefont{Lancichinetti et~al.}(2008)\citenamefont{Lancichinetti,
  Fortunato, and Radicchi}}]{lancichinetti2008benchmark}
\bibinfo{author}{\bibfnamefont{A.}~\bibnamefont{Lancichinetti}},
  \bibinfo{author}{\bibfnamefont{S.}~\bibnamefont{Fortunato}},
  \bibnamefont{and} \bibinfo{author}{\bibfnamefont{F.}~\bibnamefont{Radicchi}},
  \bibinfo{journal}{Phys. Rev. E} \textbf{\bibinfo{volume}{78}},
  \bibinfo{pages}{046110} (\bibinfo{year}{2008}).

\bibitem[{\citenamefont{Lancichinetti and
  Fortunato}(2009)}]{lancichinetti2009community}
\bibinfo{author}{\bibfnamefont{A.}~\bibnamefont{Lancichinetti}}
  \bibnamefont{and}
  \bibinfo{author}{\bibfnamefont{S.}~\bibnamefont{Fortunato}},
  \bibinfo{journal}{Phys. Rev. E} \textbf{\bibinfo{volume}{80}},
  \bibinfo{pages}{056117} (\bibinfo{year}{2009}).

\bibitem[{\citenamefont{Abbe et~al.}(2016)\citenamefont{Abbe, Bandeira, and
  Hall}}]{abbe2016exact}
\bibinfo{author}{\bibfnamefont{E.}~\bibnamefont{Abbe}},
  \bibinfo{author}{\bibfnamefont{A.~S.} \bibnamefont{Bandeira}},
  \bibnamefont{and} \bibinfo{author}{\bibfnamefont{G.}~\bibnamefont{Hall}},
  \bibinfo{journal}{IEEE Trans. Inf. Theory} \textbf{\bibinfo{volume}{62}},
  \bibinfo{pages}{471} (\bibinfo{year}{2016}).

\bibitem[{\citenamefont{Abbe and Sandon}(2015)}]{abbe2015community}
\bibinfo{author}{\bibfnamefont{E.}~\bibnamefont{Abbe}} \bibnamefont{and}
  \bibinfo{author}{\bibfnamefont{C.}~\bibnamefont{Sandon}}, in
  \emph{\bibinfo{booktitle}{Foundations of Computer Science (FOCS), 2015 IEEE
  56th Annual Symposium on}} (\bibinfo{organization}{IEEE},
  \bibinfo{year}{2015}), pp. \bibinfo{pages}{670--688}.

\bibitem[{\citenamefont{Abbe}(2017)}]{abbe2017community}
\bibinfo{author}{\bibfnamefont{E.}~\bibnamefont{Abbe}}, \bibinfo{journal}{arXiv
  preprint arXiv:1703.10146}  (\bibinfo{year}{2017}).

\bibitem[{\citenamefont{Mossel et~al.}(2013)\citenamefont{Mossel, Neeman, and
  Sly}}]{mossel2013proof}
\bibinfo{author}{\bibfnamefont{E.}~\bibnamefont{Mossel}},
  \bibinfo{author}{\bibfnamefont{J.}~\bibnamefont{Neeman}}, \bibnamefont{and}
  \bibinfo{author}{\bibfnamefont{A.}~\bibnamefont{Sly}},
  \bibinfo{journal}{arXiv preprint arXiv:1311.4115}  (\bibinfo{year}{2013}).

\bibitem[{\citenamefont{Shannon}(1948)}]{shannon}
\bibinfo{author}{\bibfnamefont{C.}~\bibnamefont{Shannon}},
  \bibinfo{journal}{Bell Syst. Tech. J.} \textbf{\bibinfo{volume}{27}},
  \bibinfo{pages}{379–423, 623–656} (\bibinfo{year}{1948}).

\bibitem[{\citenamefont{MacKay}(2003)}]{mackay2003information}
\bibinfo{author}{\bibfnamefont{D.~J.} \bibnamefont{MacKay}},
  \emph{\bibinfo{title}{Information theory, inference and learning algorithms}}
  (\bibinfo{publisher}{Cambridge university press}, \bibinfo{year}{2003}).

\bibitem[{\citenamefont{Gallager}(1962)}]{gallager1962low}
\bibinfo{author}{\bibfnamefont{R.}~\bibnamefont{Gallager}},
  \bibinfo{journal}{IRE Trans. Inf. Theory} \textbf{\bibinfo{volume}{8}},
  \bibinfo{pages}{21} (\bibinfo{year}{1962}).

\bibitem[{\citenamefont{MacKay and Neal}(1996)}]{mackay1996near}
\bibinfo{author}{\bibfnamefont{D.~J.} \bibnamefont{MacKay}} \bibnamefont{and}
  \bibinfo{author}{\bibfnamefont{R.~M.} \bibnamefont{Neal}},
  \bibinfo{journal}{Electron. Lett.} \textbf{\bibinfo{volume}{32}},
  \bibinfo{pages}{1645} (\bibinfo{year}{1996}).

\bibitem[{\citenamefont{Bierbrauer}(2005)}]{bierbrauer2005coding}
\bibinfo{author}{\bibfnamefont{J.}~\bibnamefont{Bierbrauer}},
  \bibinfo{journal}{Chapmpan \& Hall/CRC, New York}  (\bibinfo{year}{2005}).

\bibitem[{\citenamefont{Decelle
  et~al.}(2011{\natexlab{b}})\citenamefont{Decelle, Krzakala, Moore, and
  Zdeborov{\'a}}}]{decelle2011asymptotic}
\bibinfo{author}{\bibfnamefont{A.}~\bibnamefont{Decelle}},
  \bibinfo{author}{\bibfnamefont{F.}~\bibnamefont{Krzakala}},
  \bibinfo{author}{\bibfnamefont{C.}~\bibnamefont{Moore}}, \bibnamefont{and}
  \bibinfo{author}{\bibfnamefont{L.}~\bibnamefont{Zdeborov{\'a}}},
  \bibinfo{journal}{Phys. Rev. E} \textbf{\bibinfo{volume}{84}},
  \bibinfo{pages}{066106} (\bibinfo{year}{2011}{\natexlab{b}}).

\bibitem[{\citenamefont{Richardson and Urbanke}(2001)}]{richardson2001capacity}
\bibinfo{author}{\bibfnamefont{T.~J.} \bibnamefont{Richardson}}
  \bibnamefont{and} \bibinfo{author}{\bibfnamefont{R.~L.}
  \bibnamefont{Urbanke}}, \bibinfo{journal}{IEEE Trans. Inf. Theory}
  \textbf{\bibinfo{volume}{47}}, \bibinfo{pages}{599} (\bibinfo{year}{2001}).

\bibitem[{\citenamefont{Abbe and Sandon}(2016)}]{abbe2016achieving}
\bibinfo{author}{\bibfnamefont{E.}~\bibnamefont{Abbe}} \bibnamefont{and}
  \bibinfo{author}{\bibfnamefont{C.}~\bibnamefont{Sandon}}, in
  \emph{\bibinfo{booktitle}{Adv. Neural Inf. Process. Syst.}}
  (\bibinfo{year}{2016}), pp. \bibinfo{pages}{1334--1342}.

\bibitem[{\citenamefont{Erdos and R{\'e}nyi}(1960)}]{erdos1960evolution}
\bibinfo{author}{\bibfnamefont{P.}~\bibnamefont{Erdos}} \bibnamefont{and}
  \bibinfo{author}{\bibfnamefont{A.}~\bibnamefont{R{\'e}nyi}},
  \bibinfo{journal}{Publ. Math. Inst. Hung. Acad. Sci}
  \textbf{\bibinfo{volume}{5}}, \bibinfo{pages}{17} (\bibinfo{year}{1960}).

\bibitem[{\citenamefont{Neri et~al.}(2008)\citenamefont{Neri, Skantzos, and
  Boll{\'e}}}]{neri2008gallager}
\bibinfo{author}{\bibfnamefont{I.}~\bibnamefont{Neri}},
  \bibinfo{author}{\bibfnamefont{N.}~\bibnamefont{Skantzos}}, \bibnamefont{and}
  \bibinfo{author}{\bibfnamefont{D.}~\bibnamefont{Boll{\'e}}},
  \bibinfo{journal}{J. Stat. Mech. Theory Exp.}
  \textbf{\bibinfo{volume}{2008}}, \bibinfo{pages}{P10018}
  (\bibinfo{year}{2008}).

\bibitem[{\citenamefont{Peel et~al.}(2017)\citenamefont{Peel, Larremore, and
  Clauset}}]{peel2017ground}
\bibinfo{author}{\bibfnamefont{L.}~\bibnamefont{Peel}},
  \bibinfo{author}{\bibfnamefont{D.~B.} \bibnamefont{Larremore}},
  \bibnamefont{and} \bibinfo{author}{\bibfnamefont{A.}~\bibnamefont{Clauset}},
  \bibinfo{journal}{Science Adv.} \textbf{\bibinfo{volume}{3}},
  \bibinfo{pages}{e1602548} (\bibinfo{year}{2017}).

\end{thebibliography}

\end{document}